\documentclass[aps,twocolumn,
superscriptaddress,
footinbib,
pra
]{revtex4-2}
\usepackage{amsmath,amssymb,bm}
\usepackage{graphicx}
\usepackage{epstopdf}
\usepackage{latexsym}
\usepackage[normalem]{ulem} 
\usepackage[caption=false]{subfig}
\usepackage[usenames,dvipsnames]{color}
\usepackage{hyperref}
\usepackage{xcolor}
\usepackage{natbib}
\usepackage{bbold}
\usepackage{verbatim}

\hypersetup{
	colorlinks,
	linkcolor={blue!40!black},
	citecolor={blue!40!black},
	urlcolor={blue!40!black}
}

\usepackage{amsthm}
\newtheorem{theorem}{Theorem}
\newtheorem*{theorem*}{Theorem}
\newtheorem{lemma}[theorem]{Lemma}
\newtheorem{corollary}[theorem]{Corollary}

\usepackage[utf8]{inputenc} 

\usepackage{newunicodechar}
\newunicodechar{ρ}{\rho}
\newunicodechar{σ}{\sigma}
\newunicodechar{λ}{\lambda}
\newunicodechar{Λ}{\Lambda}
\newunicodechar{μ}{\mu}
\newunicodechar{ν}{\nu}
\newunicodechar{ψ}{\psi}
\newunicodechar{ϕ}{\phi}
\newunicodechar{φ}{\varphi}
\newunicodechar{π}{\pi}

\usepackage{xparse, xspace, mathtools, braket}
\usepackage{csquotes}

\let\bkbraket\braket

\let\bkket\ket
\usepackage{physics}

\RenewDocumentCommand{\matrixel}{m m m}{\bkbraket{#1|#2|#3}}

\renewcommand{\tr}{\Tr}
\DeclarePairedDelimiter\autobracket{(}{)}
\newcommand{\br}[1]{\autobracket*{#1}}
\newcommand{\complex}{\mathbb{C}}

\newcommand{\HS}{\mathcal{H}}
\NewDocumentCommand{\BO}{o}{\mathcal{B}\br{\IfValueTF{#1}{#1}{\HS}}}
\NewDocumentCommand{\DM}{o}{\mathcal{D}\br{\IfValueTF{#1}{#1}{\HS}}}

\NewDocumentCommand{\supp}{m}{\textrm{supp}\br{#1}}

\NewDocumentCommand{\PP}{o m}{\ensuremath{\IfValueTF{#1}{P_{#1}(#2)}{P(#2)}}}
\DeclarePairedDelimiterX{\infdivx}[2]{(}{)}{
	#1\delimsize|\delimsize|#2
}
\newcommand{\kld}[2]{\ensuremath{D\infdivx*{#1}{#2}}\xspace}
\NewDocumentCommand{\ent}{d()g}{\ensuremath{H\br{\IfValueTF{#1}{#1}{#2}}}}
\NewDocumentCommand{\qent}{d()g}{\ensuremath{H\br{\IfValueTF{#1}{#1}{#2}}}}

\NewDocumentCommand{\es}{m o}{
	\IfNoValueTF{#2}{
		\mathbb{\uppercase{#1}}^{\lowercase{#1}}
	}{
		\mathbb{\uppercase{#1}}^{#2}
	}
}
\newcommand{\esbraket}[2]{\bra{\es{#1}}\ket{\es{#2}}}
\newcommand{\esketbra}[2]{\ensuremath{\ketbra{\es{#1}}{\es{#2}}}}
\newcommand{\xeket}{\ket{\es{x}}}

\newcommand{\yeket}{\ket{\es{y}}}

\newcommand{\zeket}{\ket{\es{z}}}
\newcommand{\zebra}{\bra{\es{z}}}

\NewDocumentCommand{\coverlapb}{m}{\ensuremath{c_{#1}}}
\NewDocumentCommand{\coverlap}{m m}{\coverlapb{#1 #2}}

\newcommand{\IdentityMatrix}{\mathbb{1}}
\DeclarePairedDelimiter{\mnorm}{\lVert}{\rVert}
\DeclarePairedDelimiter{\mabs}{\lvert}{\lvert}

\newcommand{\gU}{\ensuremath{\textrm{U}}}
\newcommand{\gSU}{\ensuremath{\textrm{SU}}}

\newcommand{\spif}{\ensuremath{S\br{\frac{\pi}{4}}}}
\newcommand{\stpif}{\ensuremath{S\br{\frac{3\pi}{4}}}}

\begin{document}

\title{Entanglement detection in quantum many-body systems using entropic uncertainty relations}

\date{\today}

\author{Bjarne Bergh}
\affiliation{Kirchhoff-Institut f\"{u}r Physik, Universit\"{a}t Heidelberg, Im Neuenheimer Feld 227, 69120 Heidelberg, Germany}
\author{Martin G\"{a}rttner}
\affiliation{Kirchhoff-Institut f\"{u}r Physik, Universit\"{a}t Heidelberg, Im Neuenheimer Feld 227, 69120 Heidelberg, Germany}
\affiliation{Physikalisches Institut, Universit\"at Heidelberg, Im Neuenheimer Feld 226, 69120 Heidelberg, Germany}
\affiliation{Institut f\"ur Theoretische Physik, Ruprecht-Karls-Universit\"at Heidelberg, Philosophenweg 16, 69120 Heidelberg, Germany}

\begin{abstract}
We study experimentally accessible lower bounds on entanglement measures based on entropic uncertainty relations. Experimentally quantifying entanglement is highly desired for applications of quantum simulation experiments to fundamental questions, e.g., in quantum statistical mechanics and condensed-matter physics. At the same time it poses a significant challenge because the evaluation of entanglement measures typically requires the full reconstruction of the quantum state, which is extremely costly in terms of measurement statistics.
We derive an improved entanglement bound for bipartite systems, which requires measuring joint probability distributions in only two different measurement settings per subsystem, and demonstrate its power by applying it to currently operational experimental setups for quantum simulation with cold atoms. 
Examining the tightness of the derived entanglement bound, we find that the set of pure states for which our relation is tight is strongly restricted. We show that for measurements in mutually unbiased bases the only pure states that saturate the bound are maximally entangled states on a subspace of the bipartite Hilbert space (this includes product states). We further show that our relation can also be employed for entanglement detection using generalized measurements, i.e., when not all measurement outcomes can be resolved individually by the detector. In addition, the impact of local conserved quantities on the detectable entanglement is discussed.
\end{abstract}

\maketitle  

\section{Introduction}
\label{sec:intro}
Entanglement plays a key role in understanding quantum many-body phenomena \cite{Amico2008}.
In equilibrium, entanglement is tightly connected to the characteristics of the phase diagram. Out of equilibrium, the generation of entanglement is key for understanding how a closed quantum system returns to a thermal state after a quench.
While quantum simulation experiments enable the emulation of quantum dynamics problems in a scalable way \cite{Georgescu2014}, the experimental quantification of entanglement remains a challenge. 
What can be accessed comparably easily are entanglement witnesses providing a means to detect entanglement and nonclassicality \cite{Guehne2009}. 
Generally applicable procedures to measure or even bound entanglement \cite{Plenio2007} have so far relied on either full density-matrix reconstruction \cite{Haeffner2005} or require some other form of measurement that is very hard to scale up to large system sizes and to apply to higher-dimensional systems \cite{Ekert2002, Jaksch2004, Daley2012,islam_measuring_2015, kaufman_quantum_2016, elben_renyi_2018, brydges_probing_2019, elben_mixed-state_2020, zhou_single-copies_2020}.

The approach we follow in this work is rooted in the connection between entanglement and quantum uncertainty principles with side information. If access to one part of a quantum system allows the accurate prediction of measurement outcomes in the other part for two incompatible measurements, this implies the existence of correlations between the parts which cannot be purely classical, and thus entanglement must be present \cite{Wiseman2006, Reid2009}. These arguments can be formulated in terms of entropic uncertainty relations \cite{guhne_entropic_2004, coles_entropic_2017}, which relate the predictability of two incompatible measurements with the coherent information $-H(A|B)$ between subsystems $A$ and $B$. Because the coherent information serves as a lower bound on distillable entanglement \cite{devetak_distillation_2005}, entropic uncertainty relations allow one to bound this entanglement quantifier using only the probability distribution of the possible outcomes of two measurements \cite{schneeloch_quantifying_2018}. This has been successfully applied to experiments with pairs of entangled photons \cite{li_experimental_2011, schneeloch_quantifying_2019}. The goal of our work is to apply entropic uncertainty relations to quantify entanglement in quantum \emph{many-body} systems in experimentally realistic settings (see also  the  companion  paper \cite{PhysRevLett.126.190503}). The main challenge to be addressed is the limited measurement choices due to experimental constraints. These render mutually unbiased measurements, which would lead to the maximal tightness of entropic entanglement bounds, almost impossible to implement. 

With the goal of making entanglement quantification possible also with a set of measurements far from mutually unbiased, we derive a refined entropic uncertainty relation, which allows us to tighten entanglement bounds compared with previously known relations \cite{coles_improved_2014}. The crucial extension is to use the measured joint probability distribution not only for extracting conditional entropies but also for increasing the  complementarity factor which determines the amount of detectable entanglement.

The key theoretical step for this is a slight change of perspective: Entropic uncertainty relations with memory quantify the uncertainty about observables $X_A$ and $Z_A$ given that we have access to an additional memory quantum system $B$.  
Previous relations express this uncertainty through classical-quantum conditional entropies $H(X_A|B) = H(X_A B) - H(B)$, where $H(X_A B)$ is the von Neumann entropy of the classical-quantum postmeasurement state $ρ_{X_AB}$ after measuring $X$ in $A$. For the experimental application of entanglement quantification, only fully classical entropies $H(X_A|X'_B)$ are accessible since both subsystems are being measured. Therefore, one needs to apply the data-processing inequality for the additional estimate $H(X_A|X'_B) \geq H(X_A|B)$. Here we directly derive an uncertainty bound for the measurable conditional entropies $H(X_A|X'_B)$ which includes the measured bipartite probability distributions into the complementarity factor and thereby improves previously known bounds \cite{coles_improved_2014}.

We demonstrate the strength of our new entanglement bound by applying it to experimental setups realized in cold atom experiments, namely, two distinguishable particles on an optical lattice \cite{bergschneider_experimental_2019}, and a spin-1 Bose-Einstein condensate in a bipartite setting \cite{kunkel_spatially_2018, lange_entanglement_2018}.
In both cases, our relation is shown to successfully witness and bound distillable entanglement, while previous similar relations fail to do so. The requirements regarding the necessary detector resolution and the scalability of the method in terms of the required measurement statistics will likely limit its range of applicability to intermediate-size systems. Nevertheless, given the rapid progress on various quantum simulation platforms that meet these requirements, we are convinced that the entanglement bounds derived here will find ample applications for answering questions about entanglement in quantum many-body systems in and out of equilibrium and in many different experimental settings far beyond those envisioned here. 

Our paper is structured as follows:
We start by introducing necessary concepts, notation and a selection of previous results on the topic in Sec.~\ref{sec:prelim}. We then show our improved entropic uncertainty relations both for projective measurements (Sec.~\ref{sec:fsdep_ur}) and more generalized measurements in POVMs (Sec.~\ref{sec:povms}). 
Subsequently, we provide insights about the tightness of our relation in Sec.~\ref{sec:tightness} and also investigate the limits of entropic uncertainty relations in contexts where the measurement set is further constrained by conserved quantities (Sec.~\ref{sec:conserved}). Finally, we introduce two theoretical models of cold-atom systems, in which the experimental preparation of entangled states has been demonstrated recently (Sec.~\ref{sec:appl}). For both setups we present numerical simulation results which demonstrate the strength of our relation.

\section{Preliminaries}
\label{sec:prelim}

In this section, for the presentation to be self-contained, we briefly introduce the most important concepts and notation. For a more detailed account of entropic uncertainty relations and their applications we refer the reader to Ref.~\cite{coles_entropic_2017}.

Throughout this paper, we employ the following notation for objects of quantum mechanics: 
We write states in Dirac's bra-ket notation, so kets $\ket{\psi} \in \HS$ are elements of the Hilbert space $\HS$, whereas bras $\bra{\phi} \in \HS^*$ are elements of the dual space. $\bkbraket{\phi|\psi} \in \mathbb{C}$ is then equivalent to the inner product. 
We denote the set of all bounded operators on $\HS$ as
\begin{equation}
\BO \coloneqq \Set{A \colon \HS \rightarrow \HS | A \textrm{ linear and bounded} } \,.
\end{equation}
If $\HS$ is finite-dimensional (which we will assume throughout) $\BO$ is just the set of all linear operators on $\HS$. 
For elements $A \in \BO$ we write the operator norm with respect to the norm of $\HS$ as $\lVert \cdot \rVert_2$. \par

We call an element $A \in \BO$ positive, and write $A \geq 0$, if 
\begin{equation}
\matrixel{\psi}{A}{\psi} \geq 0 \quad \forall \bkket{\psi} \in \HS\,.
\end{equation} 
Note that this definition (on complex vector spaces) implies that $A$ is Hermitian. \par
We further define
\begin{equation}
\DM \coloneqq \Set{\rho \in \BO| \tr \rho = 1,\, \rho \textrm{ positive} }\,,
\end{equation}
the set of density matrices on $\HS$.

\subsection{Von-Neumann Entropy}

Let $\HS$ be a $d$-dimensional Hilbert space and $\rho \in \DM$ a density matrix. The von-Neumann entropy of $\rho$ is defined as \cite{neumann_mathematische_1971}
\begin{equation}
\qent{\rho} \coloneqq - \tr[\rho \log(\rho)] \,,
\end{equation}
where $\log$ is the matrix logarithm. Throughout this paper, $\log = \log_2$ refers to the logarithm to the base $2$. 
Diagonalizing $\rho$ as $\rho = \sum_k \lambda_k \ketbra{k}{k}$, the von-Neumann entropy becomes
\begin{equation} \label{intro:entropy_diag}
\qent{\rho} = - \sum_k \lambda_k \log \lambda_k
\end{equation}
with the convention $0 \cdot \log(0) = 0$. 
Equation \eqref{intro:entropy_diag} can also be seen as the Shannon entropy of the eigenvalues of $\rho$: 
For a discrete random variable $X$ distributed on a set of outcomes $\Omega$ according to the probability distribution $P_X \colon \Omega \rightarrow [0,1]$ with $\sum_{x\in\Omega} \PP[X]{x} = 1$, its Shannon entropy is given by
\begin{equation}\label{d:shannon_entropy}
\ent{X} = - \sum_{x \in \Omega}\PP[X]{x} \log \PP[X]{x} \,.
\end{equation}

\subsection{Quantum Relative Entropy}

For two density matrices $\rho$ and $\sigma$, the quantum relative entropy is defined as
\begin{equation}
\kld{\rho}{\sigma} \coloneqq \tr[\rho\log(\rho)] - \tr[\rho \log(\sigma)] \,,
\end{equation}
where we set $\kld{\rho}{\sigma} = \infty$ if $\supp{\rho} \not\subset \supp{\sigma}$. 
The quantum relative entropy has a number of useful mathematical properties \cite{nielsen_quantum_2010}. Below we will make use of the following:
\begin{itemize}
	\item Positivity: $\kld{\rho}{\sigma} \geq 0 \quad \forall \rho,\sigma \in \DM$
	\item Joint convexity: For $\lambda \in [0,1]$ and $\rho_1, \rho_2, \sigma_1, \sigma_2 \in \DM$:
	\begin{multline}\label{r:quantum_relative_entropy_joint_convexity}
		\kld{\lambda \rho_1 + (1 - \lambda)\rho_2}{\lambda \sigma_1 + (1 - \lambda)\sigma_2} \\ \leq  \lambda\kld{\rho_1}{\sigma_1} + (1 - \lambda)\kld{\rho_2}{\sigma_2}
	\end{multline}
	\item Monotonicity: For any quantum channel $\Lambda$ (i.e., a completely positive trace preserving linear map on bounded operators) and states $\rho, \sigma \in \DM$:
	\begin{equation}
	\label{eq:DPI}
	    \kld{\rho}{\sigma} \geq \kld{\Lambda[\rho]}{\Lambda[\sigma]}
	\end{equation}
\end{itemize}

The monotonicity property \eqref{eq:DPI} states that the relative entropy between two quantum states is nonincreasing under the application of quantum channels and is often referred to as the data-processing inequality (DPI). Quantum channels are a very large class of operations on density matrices including (generalized) measurements, unitary time evolution and interactions with an environment. For a general introduction we refer to Ref.~\cite{nielsen_quantum_2010}.

\subsection{Bipartite Systems and Entanglement}

Bipartite quantum systems are systems, for which the Hilbert space can be decomposed into a tensor product $\HS = \HS_A \otimes \HS_B$. They form the quantum analog of bivariate probability distributions. 
For such systems we can define additional entropic quantities, the quantum conditional entropy,
\begin{equation}
\qent(A|B) \coloneqq \qent{ρ_{AB}} - \qent{ρ_B} \,,
\end{equation}
and the quantum mutual information,
\begin{equation}
I(A:B) \coloneqq \qent{ρ_A} + \qent{ρ_B} - \qent{ρ_{AB}} \,.
\end{equation}
The quantum mutual information is always positive \cite{nielsen_quantum_2010}, whereas the quantum conditional entropy $\qent(A|B)$ can also be negative.

Using the joint convexity of the quantum relative entropy, one can show that the quantum conditional entropy $\qent(A|B)$ is positive for separable states and thus an entanglement witness. Moreover, its negativity is a bound on distillable entanglement, defined as the number of Bell pairs that can be extracted from an asymptotically large number of copies \cite{devetak_distillation_2005,horodecki_quantum_2009}. Thus, it constitutes a bound on an operationally relevant entanglement measure.

\subsection{Entropic Uncertainty Relations}

\subsubsection{Maassen-Uffink Relation}

One of the key properties of our description of quantum mechanics is the concept of intrinsic uncertainty. Quantum mechanics works in a way that there are certain pairs of observables which cannot both have deterministic (or arbitrarily localized) measurement statistics, irrespective of the state. This is usually demonstrated by the Robertson uncertainty relation,
\begin{equation}\label{r:robertson_ur}
σ_X\, σ_Z \geq \frac{1}{2} \abs{\ev{[\hat{X}, \hat{Z}]}} \,,
\end{equation}
where the uncertainty is quantified by the standard deviation $σ_X$ ($σ_Z$) for measuring an observable $X$ ($Z$), respectively. Such uncertainty relations can also be formulated with the entropy as an uncertainty quantifier.
The first entropic uncertainty relation that is applicable to a wide range of measurements on finite-dimensional Hilbert spaces was given in 
\textcite{maassen_generalized_1988}: 
\begin{theorem*}[Maassen and Uffink 1988]
	Let $\rho$ be a quantum state and $X$ and $Z$ be two measurements in orthonormal bases  $\{\xeket\}$ and $\{\zeket\}$, respectively. Let $H(X)$ and $H(Z)$ denote the classical Shannon-entropy of the probability distribution of measuring the state in these bases, and 
	\begin{equation} \label{d:qMU}
	q_{MU} \coloneqq - \log(\max_{x, z} \abs{\esbraket{z}{x}}^2 )\,,
	\end{equation}
	then
	\begin{equation}\label{r:eur_onepartite_MU}
	H(X) + H(Z) \geq q_{MU} + H(\rho) \,.
	\end{equation}
\end{theorem*}
Since its original discovery, various proofs of this relation have been found. A particularly nice one can be found in the appendix of \textcite{coles_entropic_2017}. 

The individual terms in the relation have the following interpretation: The sum of the entropies corresponds to the product of variances in Eq. (\ref{r:robertson_ur}), and the complementarity factor $q_{MU}$, the logarithm of the maximum overlap, quantifies the degree of incompatibility of the two measurements. If the original state is mixed [$\qent(ρ) > 0$] both our measurements will become less deterministic and we can add this term on the right-hand side. 

With $d$ the dimension of our Hilbert space, we have
\begin{equation}
\max_{x, z} \abs{\esbraket{z}{x}}^2 \geq \frac{1}{d}
\end{equation}
and two bases (or measurements with such eigenbases) that fulfill
\begin{equation}
\abs{\esbraket{z}{x}}^2 = \frac{1}{d} = \mathrm{const.}
\end{equation}
are called mutually unbiased bases (MUBs).

\subsubsection{Bipartite Uncertainty Relations}
The entropic uncertainty relation (\ref{r:eur_onepartite_MU}) holds only if there is no form of side information available that allows us to (partially) predict measurement outcomes and thus reduce uncertainty. 
One typical example of such side information occurs when the measured system $A$ is entangled with another memory system $B$ which can be measured at will. 
In that case, correlations between the two systems can be employed to reduce uncertainty, and thus relation (\ref{r:eur_onepartite_MU}) needs to be modified. 

For a subsystem $A$, possibly entangled to a memory system $B$, \textcite{berta_uncertainty_2010} showed the following entropic uncertainty relation:

\begin{theorem*}[Berta et al.\ 2010]
Let $\rho_{AB}$ be a quantum state and $X$ and $Z$ be two measurements in orthonormal bases $\{\ket{\mathbb{X}^x}_A\}$ and $\{\ket{\mathbb{Z}^z}_A\}$ on the subsystem A. Let
\begin{equation}
ρ_{XB} = \sum_x \esketbra{x}{x}_A \otimes \Tr_A\br{\esketbra{x}{x}_A ρ_{AB}}
\end{equation}
be the classical-quantum state after measuring $X$ in $A$. 
Then $H(X_A|B) = H(ρ_{XB}) - H(ρ_B)$ and similar for $Z$. Now the following relation holds:
\begin{equation} \qent(X_A|B) + \qent(Z_A|B) \geq q_{MU} + \qent(A|B) \label{r:eur_bipartite_mu} \end{equation}
with $q_{MU}$ defined as in (\ref{d:qMU}). 
\end{theorem*}

A proof can again be found in the appendix of Ref.~\cite{coles_entropic_2017} and also in Ref.~\cite{coles_uncertainty_2012}.

In contrast with the relation for a system without memory (\ref{r:eur_onepartite_MU}), now we bound not the entropy of the measurements, but the conditional entropy, i.e., the uncertainty after taking into account side information from $B$. 
This conditional uncertainty is shown to be larger than the complementarity factor $q_{MU}$ plus the quantum conditional entropy, which captures the entanglement between the two subsystems.
If $A$ and $B$ are strongly entangled, often $\qent(A|B) < 0$ and the conditional uncertainty of both $X$ and $Z$ can become arbitrarily small. If the state is separable, however, then $H(A|B) \geq 0$ and so even when conditioning on $B$, some uncertainty always remains.

\subsubsection{Bounding Entanglement}

As mentioned previously, $-H(A|B)$ is an entanglement witness and a lower bound on distillable entanglement. 
Equation (\ref{r:eur_bipartite_mu}) now gives a lower bound on $-H(A|B)$:
\begin{equation} 
\label{eq:HABequ}
- H(A|B) \geq q_{MU} -  H(X_A|B) - H(Z_A|B) \,.
\end{equation}
Unfortunately, the classical-quantum entropies $H(X_A|B)$ and $H(Z_B|B)$ cannot be measured directly. However, an upper bound
\begin{equation}
H(X_A|B) \leq \inf_{X'_B} H(X_A|X'_B) \leq H(X_A|X_B) 
\label{intro:DPI_quantumclassical_steering}
\end{equation}
can be obtained by applying the data processing inequality with a measurement in an arbitrary basis $X'_B$.
Combining equations \eqref{eq:HABequ} and \eqref{intro:DPI_quantumclassical_steering}, we have
\begin{equation}\label{intro:general_eur_bound}
- H(A|B) \geq q_{MU} -  H(X_A|X'_B) - H(Z_A|Z'_B) \,,
\end{equation}
where $q_{MU}$ can be calculated and $H(X_A|X'_B) + H(Z_A|Z'_B)$ quantifies experimentally measurable correlations. This equation summarizes the basic idea behind entanglement quantification with entropic uncertainty relations. We want to obtain a bound on an entanglement quantifier by extracting correlations between measurements in two different settings.

\subsubsection{Uncertainty Relations for POVMs}
\label{intro:povms}

Many measurements done in practice do not fall into the category of measurements in orthonormal bases, but have to be described through positive operator valued measures (POVMs). Uncertainty relations similar to (\ref{r:eur_bipartite_mu}) also exists for such generalized measurements. 

A POVM is a set of positive operators $E_k \in \BO,\ k = 1, \dots, K$ that sum to the identity: $\sum_k E_k = \IdentityMatrix$. Measuring a state $ρ$ in this POVM is then understood as obtaining outcome $k$ with probability $\tr(E_k ρ)$.
Let $\es{x}$ and $\es{z}$ be the POVM-operators for two POVMs $X$ and $Z$ on subsystem $A$. 
An entropic uncertainty relation that is analogous to those using measurements in bases was proven by \textcite{frank_extended_2013}:
\begin{subequations}
\label{r:eur_povm_frank_lieb}
\begin{align}
\qent{X_A|B} + \qent{Z_A|B} &\geq - \log(c_{FL}) + \qent{A|B} \\ c_{FL} &\coloneqq \max_{x, z} \tr(\es{x} \es{z}) \,.
\end{align}
\end{subequations}
Although the complementarity factor $-\log(c_{FL})$ reduces to (\ref{d:qMU}) for the case of measuring in bases it is a rather weak bound in general. Specifically, it is significantly weaker than factors that can be achieved for POVM uncertainty relations on only a single system without memory \cite{coles_entropic_2017}. 
\textcite{tomamichel_framework_2012} and \textcite{coles_improved_2014} proved the following alternative relation that uses a stronger overlap factor at the cost of adding additional entropy terms on the right-hand side: 
\begin{equation} \label{r:eur_povm_tomamichel}
H(X_A|B) + H(Z_A|B) \geq - \log(c_T) + H(A|B) - H(A|XB)_{\tilde{ρ}}
\end{equation}
with 
\begin{equation}
c_T \coloneqq \max_x \, \mnorm[\bigg]{\sum_z \es{x}\es{z}\es{x}}_2 \leq \max_{x,z} \norm{\sqrt{\es{x}}\sqrt{\es{z}}}^2_2
\end{equation}
and the postmeasurement state
\begin{equation} 
\tilde{ρ}_{XAB} \coloneqq \sum_x \ketbra{x}{x}_X \otimes \br{\es{x}_A \otimes \IdentityMatrix_B}ρ_{AB}\br{\es{x}_A \otimes \IdentityMatrix_B} \,.
\end{equation}

\subsubsection{State-Dependent Complementarity Factors}

One of the big issues of the previously shown entropic uncertainty relations is that they are tight only for measurements in MUBs. This can be seen easily from the proof in Ref.~\cite{coles_entropic_2017} which involves an application of the data-processing-inequality and additionally an estimate 
\begin{equation}
    \mabs{\bkbraket{\mathbb{X}^{x'}|\mathbb{Z}^{z'}}}^2 \leq \max_{x,z} \abs{\esbraket{x}{z}}^2 \,.
\end{equation}
If the measurement pair is far from mutually unbiased, this estimate will be far from tight and the relation of very little use for entanglement quantification. If the measurement pair is not mutually unbiased, the two measurements are not maximally complementary for all states, so any reasonably tight uncertainty relation will likely require some information about the occupation of problematic states in its complementarity factor. 
A first step in this direction has been taken in Ref.~\cite{coles_improved_2014} where a relation with $q$ also depending on the marginal distributions $\PP[X]{x_A}$ or $\PP[Z]{z_A}$ has been shown:
\begin{subequations}
\label{r:eur_coles_improved}
\begin{align}
H(X_A|B) + H(Z_A|B) \geq H(A|B) + q_C \\
q_C \coloneqq -\sum_x \PP[X]{x_A} \log(\max_z \coverlap{x}{z})
\end{align}
\end{subequations}
with $\coverlap{x}{z} = \abs{\esbraket{x}{z}}^2$, and one can swap $X$ and $Z$ to get a potentially better relation. 

This relation is already a significant improvement in many cases, yet there are still various pairs of measurements far from mutually unbiased for which the row-maxima $\max_z \coverlap{x}{z}$ are very close to the overall maximum $\max_{x,z} \coverlap{x}{z}$, but  $\coverlap{x}{z}$ fluctuates strongly within these rows. In these cases, (\ref{r:eur_coles_improved}) does not lead to a significant improvement.

Subsequently we show, that, if we are only interested in an uncertainty relation conditioned on measurement outcomes in $B$, we can further improve on this result and find an uncertainty relation that eliminates any maximization in $\coverlap{x}{z}$ and still only uses measurable quantities. 

\section{Fully-State-Dependent Uncertainty Relation for Projective Measurements} \label{sec:fsdep_ur}

We are now ready to state the first main result of this work, which is a more state-dependent entropic uncertainty relation for bipartite systems, which gives strictly stronger entanglement bounds than previously known relations. We call it fully state dependent, because it avoids any maximization while calculating the complementarity factor from the individual overlap elements $c_{xz}$, but instead uses maximal available measured information about the state. 

\begin{theorem}[Bipartite  State-Dependent  Uncertainty Relation] \label{fully_state_dep_bound}
	Let $\HS = \HS_A \otimes \HS_B$ be a bipartite Hilbert space. Let $X$ and $Z$ be two measurements in the ONBs $\{\xeket_A\}$ and $\{\zeket_A\}$ on $\HS_A$, and $Y$ be a measurement in the ONB $\{\yeket_B\}$ on $\HS_B$. Let $\coverlap{x}{z} = \abs{\esbraket{z}{x}}^2$. Then
	\begin{multline}
	    	H(X_A| Y_B) + H(Z_A | B) \geq H(A|B) \\ - \sum_{x, y} \PP[XY]{x_A, y_B} \log(\sum_z \coverlap{x}{z} \PP[ZY]{z_A| y_B})
	    	\label{r:eur_bipartite_fsdep}
	\end{multline}
	with $\PP[XY]{x_A, y_B} = P(X_A = x_A, Y_B = y_B)$, and the conditional distribution $\PP[ZY]{z_A| y_B} = P(Z_A = z_A|Y_B = y_B)$. 
	\begin{proof}
	    This is a special case of a similar relation for POVMs shown further below, but it is very instructive to look at its proof independently. It is inspired largely by the proofs of  less state-dependent relations in Refs.~\cite{coles_entropic_2017, coles_improved_2014, coles_uncertainty_2012}. \\
		First, notice that
		\begin{equation}
		\begin{split}
		&H(Z_A|B) - H(A|B) \\
		&= H(\rho_{ZB}) - H(\rho_B)  - H(\rho_{AB}) + H(\rho_B) \\
						&= H(\rho_{ZB}) - H(\rho_{AB}) \\
						&= \kld{\rho_{AB}}{\rho_{ZB}} \,.
		\end{split}
		\end{equation}
		Now, let 
		\begin{equation}
		    \Pi_{xy} \coloneqq \esketbra{x}{x} \otimes \esketbra{y}{y}
		\end{equation}
		be the projector onto the eigenstates of the measurement outcome $(x,y)$. 
		Define the channel $\Lambda$ that measures $X$ in $A$ and $Y$ in $B$:
		\begin{equation}
		\Lambda(\rho_{AB}) = \sum_{x, y} \Pi_{xy} \rho_{AB} \Pi_{xy} \,.
		\end{equation}
		Using the data-processing inequality, defining $\rho_{XY}\coloneqq \Lambda(\rho_{AB})$, we obtain
		\begin{align}
		&\kld{\rho_{AB}}{\rho_{ZB}} \geq \kld{\Lambda(\rho_{AB})}{\Lambda(\rho_{ZB})} \notag \\
		&=  \kld{\rho_{XY}}{\sum_{x, y, z} \Pi_{xy} \esketbra{z}{z} {\rho_{AB}}{\esketbra{z}{z}} \Pi_{xy}}  \\
		&=  \kld{\rho_{XY}}{\sum_{x, y, z} \Pi_{xy}  \coverlap{x}{z} \big(\zebra \otimes \matrixel{\es{y}}{\rho_{AB}}{\es{z}} \otimes \yeket \big)} \notag\\
		&=  \kld{\rho_{XY}}{\sum_{x, y, z}  \Pi_{xy} \coverlap{x}{z} \PP[ZY]{z_A, y_B} } \notag\,. 
		\end{align}
		Now both parts of the quantum relative entropy are diagonal in the $XY$-basis, so we can easily evaluate it to
		\begin{multline}
		\kld{\rho_{XY}}{\sum_{x, y, z}  \Pi_{xy} \coverlap{x}{z} \PP[ZY]{z_A, y_B} }   = -H(\rho_{XY}) \\- \sum_{x, y} \PP[XY]{x_A, y_B} \log(\sum_z \coverlap{x}{z} \PP[ZY]{z_A, y_B}) \,.
		\end{multline}
		With $H(\rho_{XY}) = H(X_A Y_B)$, the relation follows after adding and sutracting $H(Y_B)$ on the right-hand side.
	\end{proof}
\end{theorem}

By now estimating $\coverlap{x}{z} \leq \max_{z'} \coverlap{x}{z'}$ and using that $\sum_z \PP[ZY]{z_A|y_B} = 1$ as well as $\sum_y \PP[XY]{x_A,y_B} = \PP[X]{x_A}$, we can recover a $B$-measured version of (\ref{r:eur_coles_improved}). Thus, for the application of bounding entanglement through measurements this relation implies (\ref{r:eur_coles_improved}) and (\ref{r:eur_bipartite_mu}). 
Note however, that while using previous relations for bounding entanglement required only experimental knowledge of the joint probability distributions $P_{XX}$ and $P_{ZZ}$, our new relation without any further estimates additionally requires the measurement of $P_{XZ}$ or $P_{ZX}$. 

\subsubsection*{Example: Measuring in the Schmidt Basis}
\label{s:theory:schmidt_basis}
Our new relation has a particularly clear interpretation if the state is pure and one of the two measurements is in the Schmidt basis of the state. In this case it is also qualitatively more powerful than the standard relation (\ref{r:eur_bipartite_mu}) for any measurement pair that is not mutually unbiased.

To see this, take a pure state $ρ_{AB}$ and choose the measurements $Z$ and $Z'$ to be measurements in its Schmidt basis. Explicitly, let
\begin{equation}
\ket{\psi}_{AB} = \sum_i \sqrt{λ_i} \ket{i\/}_A \otimes \ket{i\/}_B
\end{equation}
be a Schmidt decomposition of the state (such a decomposition always exists) and then take $Z$ to be the measurement on $A$ in the ONB $\{\ket{i}_A\}$ and $Z'$ to be the measurement on $B$ in $\{\ket{i}_B\}$. This implies
\begin{equation}
ρ_{AB} = \sum_{i, j} \sqrt{λ_i λ_j} \ketbra{ii}{jj\/}\!, \quad ρ_{ZZ'} = \sum_i \lambda_i \ketbra{ii}{ii}
\end{equation}
and 
\begin{equation}
\begin{split}
H(A) &= H(B) = H(Z_A) = H(Z'_B) \\
&= H(Z_A Z'_B) = H(\{λ_i\})
\end{split}
\end{equation}
which leads to $H(Z_A|Z'_B) = 0$. Also, we have $\PP[ZZ']{i, j} = \lambda_{i\/} \delta_{ij}$.

Now, let $U = U_A\otimes U_B$ be the unitary rotation that transforms between the measurements $(X_A, Y_B)$ and $(Z_A, Z'_B)$. We will take $U_A$ = $U_B$, and write its matrix elements as $u_{ij} \coloneqq \bra{i} U_A \ket{j}$. Then, we can calculate $P_{ZY}$ as 
\begin{equation}
\PP[ZY]{i,y} = \abs{\bra{\psi}\br{\bkket{i}_A\! \otimes U_B \ket{y}_B}}^2 = \abs{\sqrt{\lambda_i\/} u_{iy\/}}^2 = \lambda_i \coverlap{i}{y} \,.
\end{equation}
Thus,
\begin{equation}
\sum_i \coverlap{i}{x} \PP[ZY]{i,y} = \sum_i \coverlap{i}{x} \coverlap{i}{y} \lambda_i \,.
\end{equation}
The key point is that this is the same as the probability distribution $P_{XY}$ of measuring the postmeasurement state $ρ_{ZZ'}$ in the $XY$ basis:
\begin{equation}
\begin{split}
 &\PP[XY]{x,y}_{ρ_{ZZ'}} \\ 
& = \bra{x}_A \otimes \bra{y}_B U^\dagger \sum_i \lambda_i \ketbra{ii}{ii} U \ket{x}_A \otimes \ket{y}_B \\
&= \sum_i \lambda_i \mabs{u_{ix\/}}^2 \mabs{u_{iy\/}}^2 = \sum_i \lambda_{i\/} \coverlap{i}{x} \coverlap{i}{y} \,.
\end{split}
\end{equation}
Now, our entanglement bound from the full-state-dependent entropic uncertainty relation reads
\begin{equation}
\begin{split}
&H(A|B) \\
&\leq H(X_A|Y_B) + \sum_{xy} P_{XY}(x_A, y_B) \log(\sum_i \coverlap{x}{i} \PP[ZY]{i_A| y_B})  \\
&= H(X_A Y_B) + \sum_{xy} P_{XY}(x_A, y_B) \log(\sum_i \coverlap{x}{i} \PP[ZY]{i_A y_B}) \\
&= -\kld{P_{XYρ_{AB}}}{P_{XYρ_{ZZ'}}} \,.
\end{split}
\end{equation}

This has the following interpretation: Measuring in the Schmidt basis shows perfect correlations between the two subsystems. Now, this alone does not demonstrate entanglement since there is also a purely classical state that shows exactly the same probability distribution, the postmeasurement state $ρ_{ZZ'}$. The entropic uncertainty relation now tells us that we can certify entanglement to the degree to which we can distinguish our real state $ρ_{AB}$ from the classical state $ρ_{ZZ'}$ by measuring in the $XY$ basis. This distinguishability is quantified by the classical relative entropy.

Note that, for pure separable states measuring in the Schmidt basis makes the relation tight. Also, we either cannot distinguish $ρ_{AB}$ from the classical state $ρ_{ZZ'}$ by measuring in the $XY$ basis, in which case we get $H(A|B) \leq 0$, or we can (possibly only to a very small degree) in which case we already demonstrate entanglement. While it is intuitively clear that for pure states any such observation demonstrates entanglement, it follows from the previously known entropic uncertainty relations only for the case of mutually unbiased measurements. 

\section{Fully-State-Dependent Uncertainty Relations For POVMs} \label{sec:povms}
The previous fully-state-dependent relation can be naturally generalized to POVMs by proving an equivalent tripartite uncertainty relation and employing the duality between tripartite and bipartite uncertainty relations. 
\subsection{Notation}
For a POVM $X$ we label the measurement operator corresponding to the measurement outcome $x$ as $\mathbb{X}^x$. 
We then implement such a POVM with $K$ elements by an isometry $V$ adding two auxiliary Hilbert spaces for our measurement result:
\begin{subequations}\label{theory:povms:notation1}
\begin{align}
V_X &\colon \HS_{AB} \rightarrow \complex^K \otimes \complex^K \otimes \HS_{AB} \\
V_X \ket{\psi_{AB}} &\coloneqq \sum_x \ket{x} \otimes \ket{x} \otimes \sqrt{\mathbb{X}^x}\ket{\psi_{AB}} \,.
\end{align}
\end{subequations}
This corresponds to adding two registers that hold the value of the measurement outcome. The measurement outcome is added twice, so that tracing out one of them removes the off-diagonal terms and acts like performing the measurement.
We write
\begin{equation}\label{theory:povms:notation2}
\rho^X \coloneqq \tilde{\rho}_{XX'AB} = V_X \rho_{AB} V^\dagger_X
\end{equation}
for the state after the isometry. The postmeasurement state is then given as $\tilde{\rho}_{XAB}$. 

\subsection{Tripartite Relations}
For many of the previous bipartite entropic uncertainty relations there exists an equivalent formulation for tripartite systems $ABC$. In this formulation, one of the two measurements is conditioned on $B$ and the other one on $C$, while the right-hand side of the relation no longer has a term for the quantum conditional entropy $\qent(A|B)$ (see Lemma \ref{lemma:bipartite_tripartite_relations} for a precise formulation). 
Tripartite uncertainty relations are often intuitively related to the monogamy of entanglement: a subsystem can show quantum correlations with one other subsystem but not with both. In that sense tripartite uncertainty relations generalize the notion of incompatibility of measurements to cases where side information is available. 

Especially in the framework of measuring in bases, the relation between bipartite and tripartite uncertainty relations has been known and used since the discovery of (\ref{r:eur_bipartite_mu}) by \textcite{berta_uncertainty_2010}. 
In the case of \enquote{coarse grained} measurements described by POVMs the correspondence is a bit more subtle, as the bipartite relation acquires additional terms from the fact that entanglement can persist in parts of the state that have not been measured. 

It appears that the best way to treat bipartite uncertainty relations for POVMs is indeed with these additional terms, for which one can then do worst-case estimates as desired. Therefore, our strategy for deriving a fully-state-dependent uncertainty relation for POVMs will be to first prove a tripartite relation, and to then obtain a bipartite relation using the following equivalence theorem.
\begin{lemma} \label{lemma:bipartite_tripartite_relations}
	Assume, that for two POVMs $X$ and $Z$ we have some tripartite uncertainty relation
	\begin{equation}\label{eur_tripartite_relation}
	H(X_A|\Lambda(B)) + H(Z_A|C) \geq q \,,
	\end{equation}
	where we allow for some measurement channel $\Lambda$ to include relations of the form $H(X_A|B)$ as well as $H(X_A|Y_B)$. $q$ can in general be state dependent, but should depend only on the reduced state $\rho_{AB}$ and not on subsystem $C$. 
	Then, this implies the bipartite uncertainty relation
	\begin{equation}\label{eur_bipartite_tripartite_relation}
	H(X_A|\Lambda(B)) + H(Z_A|B) \geq q + H(A|B) - H(A|ZB)_{\rho^{Z}}
	\end{equation}
	with $\rho^Z$ similar to (\ref{theory:povms:notation2}) and (\ref{theory:povms:notation1}).
	\begin{proof}
	    We give a simple proof of what we require for the sake of completeness. For a more general statement of this equivalence, see e.g., Ref.~\cite{tomamichel_framework_2012}.
		Let $\rho_{ABC} = \ketbra{\psi}{\psi}$ be a purification of $\rho_{AB}$, and $\rho^Z = \tilde{\rho}_{ZZ'ABC} = V_Z \rho_{ABC} V^\dagger_Z$. All subsequent entropies apply to $\rho^Z$ unless otherwise denoted.
		For any bipartite splitting of the systems $ZZ'ABC$ the two subsystems will have equal entropy by Schmidt decomposition (since $\tilde{\rho}_{ZZ'ABC}$ is pure). In particular, $H(ZZ'AB) = H(C)$  and $H(ZC) = H(Z'AB)$. Thus,
		\begin{equation}
		\begin{split}
		&H(Z|C) = H(ZC) - H(C) \\
		&= - H(ZZ'AB) + H(Z'AB) = - H(Z|Z'AB) \\
		&= - H(ZZ'A|B) + H(Z'A|B) \\
		&= - H(ZZ'A|B) + H(ZA|B) \\
		&= - H(ZZ'A|B) + H(A|ZB) + H(Z|B) \,,
		\end{split}
		\end{equation}
		where we made repeated use of the chain rule $H(AB|C) = H(A|BC) + H(B|C)$.
		Now, note that $H(ZZ'A|B)_{\rho^Z} = H(A|B)_\rho$, with $\rho = \rho_{AB}$, and thus
		\begin{equation}
		H(Z|C) = H(Z|B) - [H(A|B)_\rho - H(A|ZB)_{\rho^Z}]\,.
		\end{equation}
	\end{proof}
\end{lemma}
If the POVM $Z$ actually is an orthonormal basis, then $\mathbb{Z}^z$ is a rank-1 projector and the postmeasurement state $\tilde{\rho}_{ZAB}$ takes the form
\begin{equation}
\tilde{\rho}_{ZAB} = \sum_z \ketbra{z}{z} \otimes \mathbb{Z}^z \otimes \tr_A(\mathbb{Z}^z \rho_{AB})  \,,
\end{equation}
so $H(A|ZB) = 0$. 
If $\mathbb{Z}^z$ are not rank-1 projectors, then $H(A|ZB)$ will in general not be zero. Since $\tilde{\rho}_{ZAB}$ is classical in the $Z$-system we can write it as
\begin{equation}
H(A|ZB) = \sum_z p(z) H(A|B)_{\sqrt{\mathbb{Z}^z} \rho_{AB}\sqrt{\mathbb{Z}^z}} \,.
\end{equation}
Thus, $H(A|ZB)$ has the interpretation of the average entanglement left in the postmeasurement state.

\subsection{Fully-State-Dependent Relation}

We now prove a tripartite entropic uncertainty relation for POVMs in order to subsequently apply Lemma \ref{lemma:bipartite_tripartite_relations} to obtain the corresponding bipartite relation.
\begin{theorem}[Tripartite State-Dependent Uncertainty Relation for POVMs]			  
	\label{theorem:improved_relation_povm_tripartite}
	For any tripartite state $\rho_{ABC}$ and POVMs $X$ and $Z$ on subsystem $A$ as well as a POVM $Y$ on subsystem $B$, 
	\begin{multline} 
	\ent(X_A|\mkern 1mu Y_B) + \ent(Z_A| C) \geq \\- \sum_{x, y} \PP[XY]{x_A, y_B} \log(\sum_z h(x, z) \PP[ZY]{z_A|y_B})
	\end{multline}
	with 
	\begin{equation}
	h(x, z) \coloneqq \norm{\sqrt{\es{z}} \es{x} \sqrt{\es{z}}}_2 = \norm{\sqrt{\mathbb{Z}^z}\sqrt{\mathbb{X}^x}}^2_2 \,.
	\end{equation}
	\begin{proof}
		As in the case of Theorem \ref{fully_state_dep_bound}, this builds on the proof of a less state-dependent relation in Ref.~\cite{coles_improved_2014}.
		We start with
		\begin{equation}
		H(Z_A|C) \geq \kld{\rho_{AB}}{\sum_z \mathbb{Z}^z \rho_{AB} \mathbb{Z}^z} \,,
		\end{equation} 
		which has been shown in Ref.~\cite{coles_information-theoretic_2011} and again differently in Ref.~ \cite{coles_improved_2014}.
		We define our measurement channel 
		\begin{equation}
		\Lambda \br{\rho_{AB}} \coloneqq \rho_{XY} =  \sum_{x, y} \ketbra{x}{x} \otimes \ketbra{y}{y} \tr(\mathbb{X}^x \otimes \mathbb{Y}^y \rho_{AB})
		\end{equation}
		and, using the DPI, obtain
		\begin{align}
		&H(Z_A|C) \geq \kld{\rho_{AB}}{\sum_z \mathbb{Z}^z \rho_{AB} \mathbb{Z}^z} \notag \\
		&\geq \kld{\rho_{XY}}{\Lambda\left(\sum_z \mathbb{Z}^z \rho_{AB} \mathbb{Z}^z\right)} \\
		&= \kld{\rho_{XY}}{ \sum_{x, y, z} \ketbra{x}{x} \otimes \ketbra{y}{y} \tr(\br{\mathbb{X}^x \otimes \mathbb{Y}^y } \mathbb{Z}^z \rho_{AB} \mathbb{Z}^z)} \,. \notag
		\end{align}
		We have
		\begin{equation}
		\begin{split}
		    &\tr(\mathbb{X}^x \otimes \mathbb{Y}^y \br{ \mathbb{Z}^z \rho_{AB} \mathbb{Z}^z}) \\
		    &= \tr(\sqrt{\mathbb{Z}^z}\mathbb{X}^x \sqrt{\mathbb{Z}^z} \left(\sqrt{\mathbb{Z}^z} \otimes \mathbb{Y}^y  \rho_{AB} \sqrt{\mathbb{Z}^z} \right)) \\
		    &\leq \mnorm{\sqrt{\mathbb{Z}^z} \mathbb{X}^x \sqrt{\mathbb{Z}^z}}_2 \tr(\mathbb{Z}^z \otimes \mathbb{Y}^y  \rho_{AB} ) \\
		    &= h(x, z)  \PP[ZY]{z_A, y_B} \,,
		\end{split}
		\end{equation}
		where we used that for positive operators $A$ and $B$, we have $(\mnorm{A}_2 \mathbb{1}- A)B \geq 0$ and thus $\tr(AB) \leq \tr(\mnorm{A}_2 B)$.
		Putting everything together, we obtain
		\begin{align}
		&H(Z_A|C) + H(XY) \\
		&\geq- \sum_{x, y} \PP[XY]{x_A, y_B} \log(\sum_z \tr(\br{\mathbb{X}^x \otimes \mathbb{Y}^y} \mathbb{Z}^z \rho_{AB} \notag \mathbb{Z}^z))  \\
		&\geq - \sum_{x, y} \PP[XY]{x_A, y_B} \log(\sum_z h(x, z)  \PP[ZY]{z_A, y_B})\, \notag.
		\end{align}
		Subtracting $H(Y)$ on both sides gives the desired relation.
	\end{proof}
\end{theorem}

\begin{corollary}[Bipartite State-Dependent Uncertainty Relation for POVMs]	\label{theorem:improved_relation_povm}
	For any bipartite state $\rho_{AB}$ and POVMs $X$ and $Z$ on subsystem $A$ as well as a POVM $Y$ on subsystem $B$: 
	\begin{multline}\label{r:eur_povm_fully_statedep}
	H(X_A | Y_B) + H(Z_A | B) \geq \\ H(A|B) - H(A|ZB)_{\rho^Z} + q_{FSDP}
	\end{multline}
	with 
	\begin{equation}
		q_{FSDP} = -\sum_{x, y} \PP[XY]{x_A, y_B} \log(\sum_z h(x, z) \PP[ZY]{z_A|y_B}) \,.
	\end{equation}
	\begin{proof}
		Apply Lemma \ref{lemma:bipartite_tripartite_relations} to Theorem \ref{theorem:improved_relation_povm_tripartite}. 
	\end{proof}
\end{corollary}

Note that the given version does not directly imply the state-independent (\ref{r:eur_povm_tomamichel}) and a marginal-dependent version similar to (\ref{r:eur_coles_improved}) for POVMs (see \textcite{coles_improved_2014} for an explicit formulation). The key issue is that these less state-dependent relations can use a complementarity factor where the matrix norm is applied to a larger sum of operators, namely
\begin{equation}
	 h(x) \coloneqq \norm{\sum_z \es{x}\es{z}\es{x}}_2 \,,
\end{equation}
while bounding by the maximum in (\ref{r:eur_povm_fully_statedep}) yields
\begin{equation}
h(x, z') \leq \max_{z} h(x,z) = \max_{z} \norm{\sqrt{\es{z}} \es{x} \sqrt{\es{z}}}_2 \,.
\end{equation}
\textcite{coles_improved_2014} showed
\begin{equation}
\norm{\sum_z \es{x}\es{z}\es{x}}_2 \leq \max_{z} \norm{\sqrt{\es{z}} \es{x} \sqrt{\es{z}}}_2
\end{equation}
with no equality in general. 

\subsection{Entanglement Witnessing}
When using the previously shown entropic uncertainty relations for POVMs instead of those for orthonormal bases, one has to deal with the issue that the entanglement quantifier they use is not just the quantum conditional entropy but the quantum conditional entropy minus the remaining quantum conditional entropy in the postmeasurement state. 
Here we show that this modified term is still an entanglement witness, i.e.\ it remains positive for separable states. 
\newcommand{\rt}{\tilde{\rho}}
\begin{theorem}[Entanglement Witness for POVMs]
	\label{t:povm_entanglement_witnessing}
	For any separable bipartite state $\rho_{AB}$ and any POVM $Z$ on subsystem $A$, it holds that
	\begin{equation}
	H(A|B) - H(A|ZB)_{\rho^Z} \geq 0 \,.
	\end{equation}	
	\begin{proof}
		Again write $\tilde{\rho}_{ZZ'AB} = V_{Z\/} \rho_{AB}V^\dagger_Z$. Then $H(A|B)_\rho = H(ZZ'A|B)_{\tilde{\rho}}$.
		For any direct product state $\rho_{AB} = \rho_{A} \otimes \rho_{B}$ we get $H(ZZ'A|B) = H(ZZ'A)$ and $H(A|ZB) = H(A|Z)$. Then, by the Araki-Lieb inequality \cite{araki_entropy_1970}, we have
		\begin{equation}
		\begin{split}
		H(ZZ'A) &\geq |H(ZA) - H(Z')| \geq H(ZA) - H(Z')\\
		&= H(ZA) - H(Z) = H(A|Z)\,.
		\end{split}
		\end{equation}

		For separable states $\rho_{AB} = \sum_k p_k \rho_{A}^k \otimes \rho_{B}^k$ we can rewrite our expression as a relative entropy and then use joint convexity. 
		We have
		\begin{equation}\label{povms:sep_statement}
		\rt_{ZZ'AB} = \sum_k p_k \rt_{ZZ'AB}^k = \sum_k p_k \rt_{ZZ'A}^k \otimes \rt_{B}^k \,,
		\end{equation} 
		which leads to
		\begin{align}
		&H(ZZ'A|B) - H(A|BZ) \notag \\
		&= H(ZZ'AB) + H(BZ) - H(ABZ) - H(B) \\
		&= -\kld{\rt_{ZZ'AB}\otimes \rt_{ZB}}{\frac{I}{d_Z} \otimes \rt_{ZAB} \otimes \frac{I}{d_Z} \otimes \rt_{B}} \notag \\ &\phantom{{}={}}+ 2 \log(d_Z)\notag \,.
		\end{align}
		Now, using the separability (\ref{povms:sep_statement}) and joint convexity of the relative entropy, we obtain
		\begin{equation}
		\begin{split}
		&H(ZZ'A|B) - H(A|BZ)\\
		&\geq \sum_{k,j} p_k p_j \br{H(Z|Z'AB)_{\rt^k} + H(Z|B)_{\rt^j}} \\
		&= \sum_k p_k \br{ H(Z|Z'AB)_{\rt^k} + H(Z|B)_{\rt^k}} \\
		&= \sum_k p_k \br{ H(ZZ'A|B)_{\rt^k} - H(A|BZ)_{\rt^k}} \geq 0 \,,
		\end{split}
		\end{equation}
		where the $\rt^k$ are direct product states, so $H(ZZ'A|B)_{\rt^k} - H(A|BZ)_{\rt^k} \geq 0$.
	\end{proof}
\end{theorem}

\section{Properties and Limitations of Entanglement Quantification with Entropic Uncertainty Relations}
\label{sec:properties}

In this section we investigate properties of our relation and of entanglement quantifiers based on entropic uncertainty relations in general. The first part of this section will deal with tightness of fully measured relations, while the second part answers what one can expect in scenarios where the set of available measurements is restricted by some conservation law. 

\subsection{Tightness}\label{sec:tightness}
For an application of entropic uncertainty relations to entanglement quantification, one needs to be aware that the classical-quantum conditional entropies $H(Z_A|B)$ cannot be measured. Experimentally, only classical-classical conditional entropies $H(Z_A|Z'_B)$ with an arbitrary measurement $Z'$ on $B$ are accessible. The data processing inequality tells us that
\begin{equation}
    \label{tightness:b-meausurement-dpi}
    H(Z_A|B) \leq H(Z_A|Z'_B)
\end{equation}
for any measurement $Z'$ on $B$. Thus, all previously stated entropic uncertainty relations imply a formulation with such classical-classical entropies. If the measurement pair $(X_A, Z_A)$ is mutually unbiased, indeed all these relations are equivalent, i.e.\ (with $d$ the dimension of $\HS_A$) 
\begin{equation}
    q_{MU} = q_{FSD} = q_{C} = \log(d) \,.
\end{equation}

In this section we investigate what effect conditioning on measurement results instead of the full quantum system $B$ has on the tightness of these relations. We will mostly be dealing with pure states and measurement pairs that are mutually unbiased. While one would expect that this should be somewhat ideal circumstances, we will see that already there the set of tight states is very limited. 

For the relation by Berta et al.\ (\ref{r:eur_bipartite_mu}) and with fixed measurements $X$ and $Z$ related by a Fourier transformation, Ref.~\cite{coles_relative_2011} tried to find all tight states, but achieved classification only with some additional restrictions.

In this section we  interest ourselves in a slightly different question: For which pure states $ρ_{AB}$ does there exist a pair of mutually unbiased bases $X_A$ and $Z_A$ of $\HS_A$ and arbitrary measurements $X'_B$ and $Z'_B$ of $\HS_B$ such that
\begin{equation} \label{tightness:measured-relation}
    H(X_A|X'_B) + H(Z_A|Z'_B) = H(A|B) + \log(d)\, .
\end{equation}
If we were to consider the relation conditioning on the quantum system instead, i.e.\ if we look for states which satisfy
\begin{equation} \label{tightness:quantum-relation}
    H(X_A|X'_B) + H(Z_A|B) = H(A|B) + \log(d) \, ,
\end{equation}
answering this question is fairly straightforward: Choose bases $X_A$ and $X'_B$ such that $ρ_{AB}$ has a Schmidt decomposition in these bases. Then $H(X_A|X'_B) = 0$. Furthermore, since $ρ_{AB}$ is pure, also $\Tr_A(\ketbra{z}{z} ρ_{AB})$ is rank one (and thus pure after pulling out the normalization), and using that $ρ_{ZB}$ is classical in A we get $H(Z_A|B) = H(Z_A) - H(B)$. Hence, the uncertainty relation reduces to $H(Z_A) \geq \log(d)$ which can only be achieved with equality. Thus, for \emph{all} pure states choosing one measurement to be the Schmidt basis makes the relation tight.

This changes drastically if we consider (\ref{tightness:measured-relation}) instead of (\ref{tightness:quantum-relation}). Our key result is the following:

\begin{theorem}[Tightness for Pure States and MUB Measurements]
\label{theorem:tightness}
    Let $ρ_{AB}$ be pure. If there exists a pair of mutually unbiased bases ($X_A$, $Z_A$) on $\HS_A$ and arbitrary measurements $X'_B$ and $Z'_B$ on $\HS_B$, such that
\begin{equation}\label{tightness:theorem-equality}
    H(X_A|X'_B) + H(Z_B|Z'_B) = H(A|B) + \log(\dim \HS_A)    
\end{equation}
then all nonzero Schmidt coefficients of $ρ_{AB}$ are equal. 
\end{theorem}

Note that all nonzero Schmidt coefficients being equal is equivalent to the following statement: There exist subspaces $\mathcal{G}_A \subset \HS_A$ and $\mathcal{G}_B \subset \HS_B$ with the same dimension such that $ρ_{AB}$ is a maximally entangled state on $\mathcal{G}_A \otimes \mathcal{G}_B$ embedded into $\HS_A \otimes \HS_B$. In the case of $\dim \mathcal{G}_A = \dim \mathcal{G}_B = 1$ this \enquote{maximally entangled state} is just a product state. 

\begin{proof}
Let $X_A$, $Z_A$ be mutually unbiased and $X'_B$, $Z'_B$ be arbitrary on $\HS_B$, such that (\ref{tightness:theorem-equality}) holds. 
It follows from the relation by Berta et al.\ (\ref{r:eur_bipartite_mu}) that for equality in (\ref{tightness:theorem-equality}) three conditions have to be met: First, the relation (\ref{r:eur_bipartite_mu}) has to be tight, and additionally for the bases $X'_B$ and $Z'_B$ of $\HS_B$ the following two conditions must be fulfilled: $H(X_A|B) = H(X_A|X'_B)$ and $H(Z_A|B) = H(Z_A|Z'_B)$ (it follows from Lemma \ref{appendix:lemma_zero_discord} in the appendix that allowing $X'_B$ and $Z'_B$ to be POVMs is not more general). We show that, if $ρ_{AB}$ has two or more distinct nonzero Schmidt coefficients then these three conditions cannot be met at the same time, which proves the claim that Eq.~\eqref{tightness:theorem-equality} implies that all Schmidt coefficients have to be equal. 

The relation (\ref{r:eur_bipartite_mu}) by Berta et al.\ can be proven by using an argument similar to our proof of Theorem \ref{fully_state_dep_bound} which invokes the data processing inequality,
\begin{equation}
\kld{ρ_{AB}}{ρ_{ZB}} \geq \kld{\Lambda[ρ_{AB}]}{\Lambda[ρ_{ZB}]} 
\end{equation}
with $\Lambda$ being the quantum channel that measures system $A$ in the $X_A$ basis. Thus, for the relation to be tight, the data processing inequality has to be tight in this specific instance.

\subsubsection{Tightness of the DPI}
For a quantum channel $\Lambda \colon \mathcal{D}(\HS) \rightarrow \mathcal{D}(\HS')$ Petz \cite{petz_sufficient_1986} showed that
\begin{equation}
\label{theory:tightness:equality}
\kld{Λ[ρ]}{Λ[σ]} = \kld{ρ}{σ} 
\end{equation}
if and only if there exists a CPTP recovery channel $\tilde{Λ}\colon \mathcal{D}(\HS') \rightarrow \mathcal{D}(\HS)$ (which may depend on $ρ$ and $σ$) that inverts $Λ$ on both $ρ$ and $σ$, i.e.
\begin{equation}
\tilde{Λ}[Λ[ρ]] = ρ, \quad \tilde{Λ}[Λ[σ]] = σ \,.
\end{equation}
Furthermore, this recovery channel can (if it exists) always be chosen of the form
\begin{equation}\label{theory:tightness:recovery_map}
\tilde{Λ}[\omega] = \sqrt{σ}Λ^*\left[\left(\sqrt{Λ[σ]}\right)^{-1} \omega \, \br{ \sqrt{Λ[σ]}}^{-1}\right]\sqrt{σ} 
\end{equation}
with $Λ^*$ the adjoint of $Λ$ with respect to the Hilbert-Schmidt inner product on $\BO$. Since $Λ$ is trace preserving, $Λ^*$ is always unital, so by construction this form ensures $\tilde{Λ}[Λ[σ]] = σ$. Note that $\kld{ρ}{σ}$ is not symmetric in $ρ$ and $σ$, so $ρ$ and $σ$ can in general not be swapped in this expression. 

Equation \eqref{theory:tightness:recovery_map} implies that the DPI is tight for the two states $ρ$ and $σ$ if and only if
\begin{equation}\label{petz_explicit_dpi_tightness_condition}
ρ = \sqrt{σ}Λ^*\left[\left(\sqrt{Λ[σ]}\right)^{-1} Λ[ρ] \, \br{ \sqrt{Λ[σ]}}^{-1}\right]\sqrt{σ} \,.
\end{equation}
Inserting $\sigma = ρ_{ZB} = \sum_z \ketbra{z}{z} \otimes ρ_{B}^{(z)}$ we find
\begin{equation}
\begin{split}
    \Lambda[ρ_{ZB}] &= \sum_{x,z} \ketbra{x}{x} \ketbra{z}{z} \ketbra{x}{x} \otimes ρ_B^{(z)} \\
    &= \frac{1}{d} \sum_x \ketbra{x}{x} \otimes \sum_z ρ_B^{(z)}\\ 
    &= \frac{1}{d} \IdentityMatrix_A \otimes \tr_A(ρ_{ZB})
\end{split}
\end{equation}
and thus the condition reduces to
\begin{equation} \label{tightness:contradiction}
ρ_{AB} = d \sqrt{ρ_{ZB}} (\IdentityMatrix_A \otimes B^{-\frac{1}{2}}_Z) ρ_{XB} (\IdentityMatrix_A \otimes B^{-\frac{1}{2}}_Z) \sqrt{ρ_{ZB}}\,,
\end{equation}
where $B^{-\frac{1}{2}}_Z = (\tr_A(ρ_{ZB}))^{-\frac{1}{2}}$ and we used that for measurement channels $\Lambda^* = \Lambda$. 

The strategy is now to show that all quantities on the right-hand side are block-diagonal in blocks corresponding to the values of Schmidt coefficients of $ρ_{AB}$, with block sizes corresponding to the degeneracy of the respective coefficient value. On the other hand, the left-hand side $ρ_{AB} = \ketbra{\psi}{\psi}$ will always also contain off-diagonal terms. Thus, if there are two or more distinct nonzero Schmidt coefficients, we get a contradiction. 

\subsubsection{Zero Quantum Discord and Schmidt Decompositions}
The statement that there exists a basis $X'_B$ such that $H(X_A|B) = H(X_A|X'_B)$ is equivalent to $ρ_{XB}$ having zero quantum discord, where quantum discord is defined as the difference $\inf_{X'_B} H(X_A|X'_B) - H(X_A|B)$. It is known \cite{modi_classical-quantum_2012, datta_condition_2011, dakic_necessary_2010, coles_entropic_2017} that a state has zero quantum discord if and only if it is classical in subsystem $B$ (a proof of this statement is given in Lemma \ref{appendix:lemma_zero_discord} in the appendix). This implies that there exists a measurement $X''$ on $B$ [this will not necessarily be any measurement $X'$ that satisfies $H(X_A|B) = H(X_A|X'_B)$] such that $\rho_{XB} = \rho_{XX''}$. Since $H(X_A|X'_B) = H(X_A|X''_B)$, also $X''$ will lead to a tight relation if $X'$ does, and thus we can assume $X' = X''$ (this is just to simplify notation). If $ρ_{AB} = \ketbra{\psi}{\psi}$ we can expand $\ket{\psi}$ in the $X_A \otimes X'_B$ basis
\begin{equation}
    \ket{\psi} = \sum_{k,l} \alpha_{kl} \ket{k}_{X_A} \otimes \ket{l}_{X'_B} \, ,
\end{equation}
and find that
\begin{equation}
    ρ_{XB} = \sum_{k, l, l'} \alpha_{kl}\alpha^*_{kl'} \ketbra{k}{k}_{X_A} \otimes \ketbra{l}{l'}_{X'_B}\, .
\end{equation}
The condition that $ρ_{XB} = ρ_{XX'}$ then gives
\begin{equation}
\forall k,l,l'\colon \, \alpha_{kl\/} \alpha_{kl'\/}^* \propto \delta_{ll'\/}  \quad \Rightarrow \quad \forall k,l\colon \, \alpha_{kl} \propto \delta_{ll_0(k)} \,,
\end{equation}
and thus $\ket{\psi}$ can be written as
\begin{equation}
    \ket{\psi} = \sum_k \alpha_{k} \ket{k}_{X_A} \otimes \ket{l_0(k)}_{X'_B}\, .
\end{equation}
Here, $l_0(k)$ can still take the same value twice for different $k$ so this is not yet a Schmidt decomposition. However, we can rearrange the sum as
\begin{equation}
    \ket{\psi} = \sum_l \lambda_l \left(\sum_k \alpha_k^{(l)} \ket{k}_{X_A}\right) \otimes  \ket{l}_{X'_B} \, ,
\end{equation}
with $\lambda_l$ and $\alpha_k^{(l)}$ defined suitably. By absorbing complex phases into a redefinition of the $X'_B$ basis we can always chose $\lambda_k$ to be real and positive.  Thus, we see that $\ket{\psi}$ has a Schmidt decomposition consisting of vectors in the $X'_B$ basis on $\HS_B$. Furthermore, since $ρ_{XB} = ρ_{XX'}$, the $B$ part of $ρ_{XB}$ is diagonal in this basis and thus $ρ_{XB}$ contains no cross-terms of different subspaces corresponding to different Schmidt coefficients.

The exact same argument can be applied to $ρ_{ZB}$ to get a different Schmidt decomposition of $\ket{\psi}$:
\begin{equation}
    \ket{\psi} = \sum_l \lambda_l \left(\sum_k \tilde{\alpha}_k^{(l)} \ket{k}_{Z_A}\right) \otimes \ket{l}_{Z'_B} \, .
\end{equation}
Now, Schmidt decompositions are not unique, so $ρ_{ZB}$ and $ρ_{XB}$ having diagonal $B$ part in two different decompositions does not imply that also their product is diagonal in either of those bases. However, the subspaces spanned by all the Schmidt vectors of a specific Schmidt coefficient value are independent of the chosen Schmidt decomposition. This corresponds to the well-known statement that the singular value decomposition is unique up to unitary rotations within the subspaces corresponding to the degeneracies of the different singular values. Thus $ρ_{XB}$ and $ρ_{ZB}$ having no off-diagonal terms from these different subspaces implies the same also for their product. Additionally, $B_Z^{-\frac{1}{2}}$ is derived from $\Tr_A(ρ_{ZB})$ and thus is also block-diagonal. This establishes the contradiction in (\ref{tightness:contradiction}) if there are two or more distinct nonzero Schmidt coefficients, and thus completes the proof of Theorem~\ref{theorem:tightness}.
\end{proof}
The necessary condition of Theorem~\ref{theorem:tightness} is not obviously sufficient. It is fairly easy to see that any product state will be tight if one chooses either $X_A$ or $Z_A$ to contain one of its product vectors. Similarly, a maximally entangled state on $\HS_A \otimes \HS_B$ will be tight even for arbitrary mutually unbiased bases $(X_A, Z_A)$ given an appropriate choice of $X'_B$ and $Z'_B$. Furthermore, a maximally entangled state on a subspace $\mathcal{G}_A \otimes \mathcal{G}_B \subset \HS_A \otimes \HS_B$, where $\dim \mathcal{G}_A$ divides $\dim \HS_A$, can be made tight by having $X_A$ contain the vectors spanning $\mathcal{G}_A$ at the right indices and choosing $Z_A$ as its Fourier transformation. However, for embedded maximally entangled states of subspaces with arbitrary dimension this is in general not correct, and it is not clear if there exists measurement choices which make the relation tight in that case.

\subsection{Limited Measurements due to Conserved Quantities}\label{sec:conserved}
In many practical applications the translation between the two measurements $X$ and $Z$ is implemented by a unitary operation generated through time evolution under some Hamiltonian. If the system has conserved quantities, the set of implementable unitaries will be limited, as all measurement operators then also have to commute with this quantity. If furthermore, the state consists of a superposition of different values of these conserved quantities, this can lead to entanglement that is undetectable through entropic uncertainty relations, because its correlations cannot be distinguished from classical through the set of implementable measurements. A typical example is local particle number, which cannot be changed by local operations, but the system can be prepared in a superposition of different particle number distributions between two subsystems \cite{Lukin2019}, e.g., through applying a beam splitter. This leads to a form of bipartite entanglement of the particle numbers in each subsystem which is undetectable with entropic uncertainty relations. 

Quantitatively, one gets the following:
\newcommand{\rb}{\overline{ρ}}
\NewDocumentCommand{\postn}{o m}{\ensuremath{#2^{(\IfNoValueTF{#1}{n}{#1})}}}
\begin{theorem}[Bipartite State-Dependent Uncertainty Relation with Local Conservation Laws] \label{theorem:conserved_quantities}
	Let $\HS = \HS_A \otimes \HS_B$. Let $N_A \otimes N_B$ be a Hermitian operator on $\HS$, and $\postn{\Pi} = \postn[n_A]{\Pi}_A \otimes \postn[n_B]{\Pi}_B$ be the projectors onto its eigenspaces enumerated by $n = (n_A, n_B)$.
	Let $X_A, Z_A$ be generalized measurements (i.e., POVMs) on $A$ such that their measurement operators $\es{x}$, $\es{z}$ both commute with $N_A$. Let similarly $X'_B$ and $Z'_B$ be generalized measurements on $B$ such that their measurement operators commute with $N_B$. Then
	\begin{multline}
	\ent(X_A|X'_B) + \ent{Z_A|Z'_B} \geq \\ q_{FSDP} + H(A|B)_{\rb} - H(Z|AB)_{\rb^Z}
	\end{multline} 
	with 
	\begin{equation}
	\rb = \sum_n \postn{\Pi} ρ_{AB\/}\, \postn{\Pi}
	\end{equation}
	and $\rb^Z$ similar to (\ref{theory:povms:notation2}). Furthermore, if $X_A$ and $Z_A$ are measurements in orthonormal bases, then
	\begin{equation}
		\ent(X_A|X'_B) + \ent{Z_A|Z'_B} \geq q_{FSD} + H(A|B)_{\rb} \,.
	\end{equation} 
	
	\begin{proof}
		Since all observables commute with the projectors $\postn{\Pi}$, in the state $\rb$ the probability of any outcome corresponding to measurement operators $\es{A}[] \otimes \es{B}[]$ is given by
		\begin{align}
			\tr(\rb\, \es{A}[] \otimes \es{B}[]) &= \tr(\sum_n \postn{\Pi} ρ\/\, \postn{\Pi} \es{A}[] \otimes \es{B}[]  ) \\
			&= \tr(\sum_n ρ\/\, \postn{\Pi} \es{A}[] \otimes \es{B}[]) =  \tr( ρ\/\, \es{A}[] \otimes \es{B}[] ) \,. \notag
		\end{align}
	Thus, all measured conditional entropies and state-dependent complementarity factors are equal for $ρ$ and $\rb$. The claim then follows from an application of (\ref{r:eur_povm_fully_statedep}) to $\rb$. 
	\end{proof}
\end{theorem}
We call $H(A|B)_{\rb}$ the configurational part of the conditional entropy $H(A|B)_ρ$, because it describes the correlations of the configurations within the conservation-law sectors. Given that the state $\rb$ is related to $ρ$ by the application of a quantum channel that measures the conserved quantity, using the data-processing inequality gives
\begin{equation}
-\qent(A|B)_{\rb} \leq -\qent(A|B)_{ρ} \,,
\end{equation}
i.e.\ the entanglement witnessed by the configurational part is always less than the total entanglement.

As an example, consider again a system where particles can fluctuate between two spatial regions which make up the two parts of our bipartite Hilbert space. We assume that the particle number within each subsystem is conserved, and the total particle number is fixed to a single value $N$. If the initial state is pure, one easily calculates \cite{Lukin2019}
\begin{equation}
-\qent(A|B)_{ρ} = \qent(B)_{ρ} = -\qent(A|B)_{\rb} + \ent(\{p(n)\}) \,,
\end{equation}
where $p(n)$ is the probability distribution of finding $n$ particles in subsystem $A$ (and thus $N - n$ in $B$). So, the true entanglement entropy separates into its configurational part and a part coming from particle number fluctuations between subsystems, which is undetectable using entropic uncertainty relations. Similar relations between $-H(A|B)_ρ$ and $-H(A|B)_{\rb}$ can easily be shown if $ρ$ is a mixture of pure states with different particle numbers, or is created from a (possibly mixed) initial state through a beamsplitter.

\section{Applications}\label{sec:appl}

We present two examples of physical systems where the set of easily accessible measurements is severely restricted and does not include MUBs. We show that for these systems, our new uncertainty relation allows us to obtain meaningful bounds on entanglement while all the previous relations with less state-dependent complementarity factors fail to do so. 
\subsection{Two Distinguishable Particles}
\label{c:impl:disting}
The first system we consider, is a simple systems of two distinguishable particles on a one-dimensional (1D) lattice. This is inspired by Ref.~\cite{bergschneider_experimental_2019}, where entanglement of such two particles on two lattice sites was quantified using density-matrix reconstruction.
We will show, that just using unitary time evolution and occupation measurements we can use entropic uncertainty relations to witness and bound (although not tightly bound) entanglement between the two particles. 

\subsubsection{Model}
The system consists of two distinguishable particles, experimentally realized using different internal states, e.g., hyperfine states of $^6$Li in Ref.~\cite{bergschneider_experimental_2019}, interacting on a 1D chain of lattice sites. 
Introducing creation and annihilation operators $a_{1, i}, a^\dagger_{1,i}, a_{2,i},  a^\dagger_{2,i}$ for each particle and lattice site $i$, we can write the Hamiltonian as
\begin{equation}\label{d:disting_part:hamiltonian}
H = -J \sum_{p=1}^2\sum_{i = 1}^{L-1} (a^\dagger_{p,i\/}\,a_{p,i+1} + \text{h.c.} ) + U \sum_{i=1}^{L} \hat{n}_{1,i\/} \, \hat{n}_{2,i}
\end{equation}
with $L$ lattice sites, hopping strength $J > 0$, interaction strength $U$, and particle number operator $\hat{n}_{p,i} \coloneqq a^\dagger_{p,i}\, a_{p,i}$. Note that we do not use periodic boundary conditions.
The system's Hilbert space is the tensor product of the Hilbert spaces of the individual particles, so we study entanglement between the two particles and not between spatially separated regions.

\subsubsection{Ground State}

\begin{figure}
	\includegraphics[width=\linewidth]{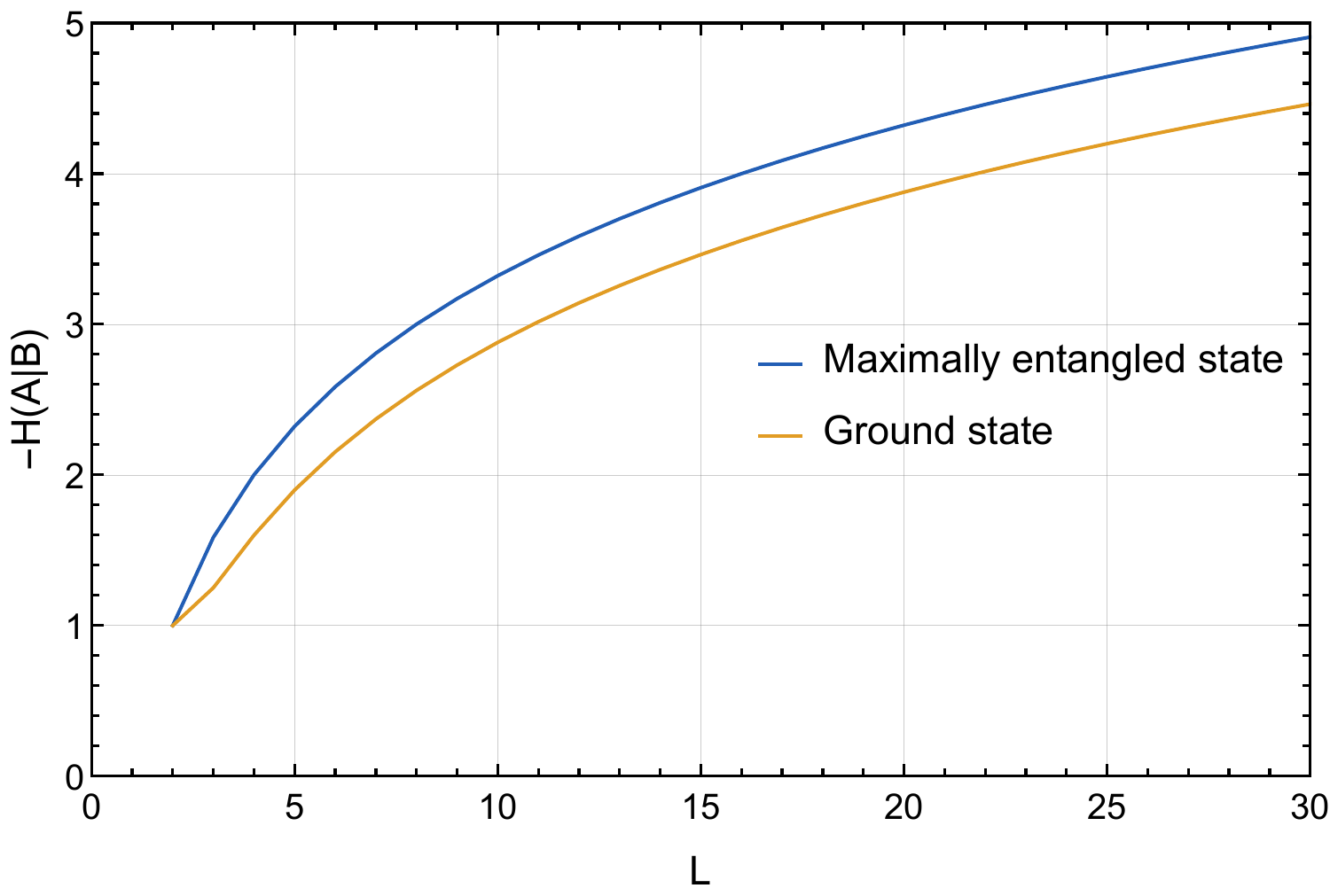}
	\caption{Entanglement entropy between two particles in the ground state of a 1D Hubbard model in the highly attractive regime ($\abs{U} \gg J$ and $U < 0$), compared with the entanglement entropy of a maximally entangled state on the same Hilbert space.}
	\label{f:disting:ent-entropy}
\end{figure}

For $\mabs{\mkern 2mu U} \gg J$ and $U < 0$, it is energetically favorable for the two particles to occupy the same lattice site, so the ground state is approximately
\begin{equation}
\ket{\psi} \approx \sum_i c_i \ket{i,i} \,,
\end{equation}
where $c_i$ are the coefficients of the single-particle ground state
\begin{equation}
\ket{\psi_1} = \sum_i c_i \ket{i} \,.
\end{equation}
Note that, since we did not employ periodic boundary conditions, the single-particle ground state is not uniform but will show decreased population towards the boundary.
An exception is the case of only two lattice sites (which was implemented in Ref.~\cite{bergschneider_experimental_2019}). Here every lattice site is on the boundary, and the bipartite ground state is a maximally entangled state. Figure \ref{f:disting:ent-entropy} shows the entanglement entropy of the two-particle ground state as a function of the number of lattice sites. 

\subsubsection{Measurement Directions}

The natural measurement in this system is the detection of the positions of both particles on the lattice. We will label the corresponding basis states as $\{\ket{i}\}_{i=1}^{L}$ for one particle, and $\{\ket{i_1, i_2}\}_{i_1,i_2=1}^{L}$ for the bipartite states, where each index $i$ corresponds to the lattice site the particle is on (site basis).  
A measurement in a different basis can be performed by letting the system evolve under a Hamiltonian with different parameters $J$ and $U$ (essentially a quantum quench) before detecting the atom positions. 
To be suitable as a second measurement in our entropic uncertainty relations, the applied time evolution must be local in the two Hilbert spaces $\HS_A$ and $\HS_B$, i.e.\ it must decompose as $e^{i tH} = R = R_A \otimes R_B$ (with $\hbar = 1$). For the Hamiltonian in (\ref{d:disting_part:hamiltonian}), this is only true if the particles are noninteracting, so $U = 0$. Thus, we remain with a one-parametric set of possible second measurement directions which are related to the occupation basis by the unitary
\begin{equation}\label{impl:disting:indep-tunneling:unitary}
R(t) = e^{i t H(J = 1,\, U = 0)} \,.
\end{equation} 
The overlap elements $c_{xz}$ are then given by the absolute values squared of the entries in this unitary matrix (\ref{impl:disting:indep-tunneling:unitary}). 
\begin{figure}
	\centering
	\includegraphics[width = \linewidth]{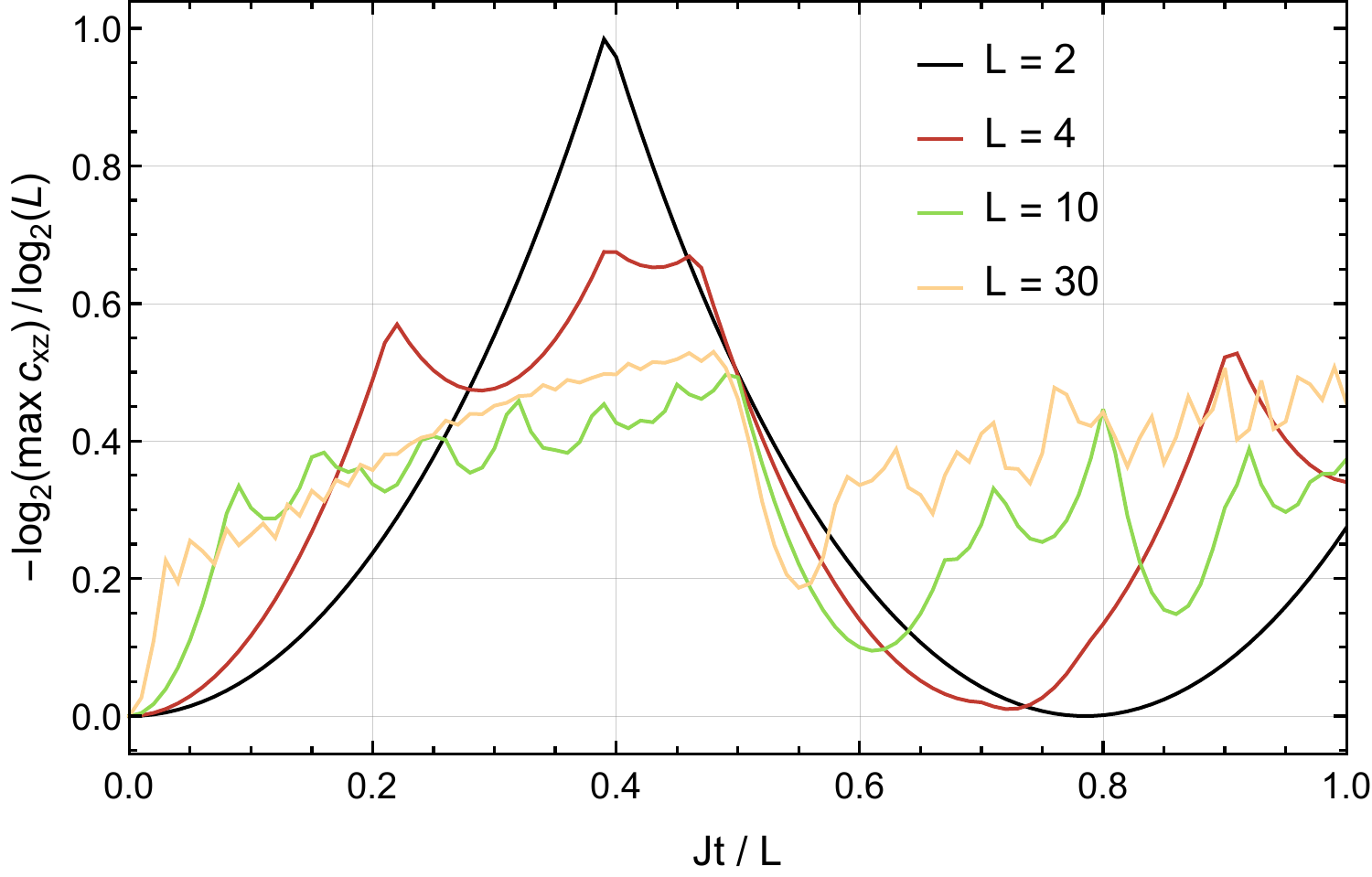}
	\caption{Normalized complementarity factor $- \log(\max_{x, z} c_{xz})$ for a measurement $X$ in the site basis and a second measurement $Z$, for varying number of lattice sites $L$. The second measurement is implemented by letting the particles evolve independently under the hopping Hamiltonian for time $t$ before measuring their positions on the lattice. The normalization is chosen such that a value of one corresponds to measuring in MUBs.}
	\label{f:disting:overlap-max}
\end{figure}

\begin{figure}
	\centering
	\includegraphics[width =\linewidth]{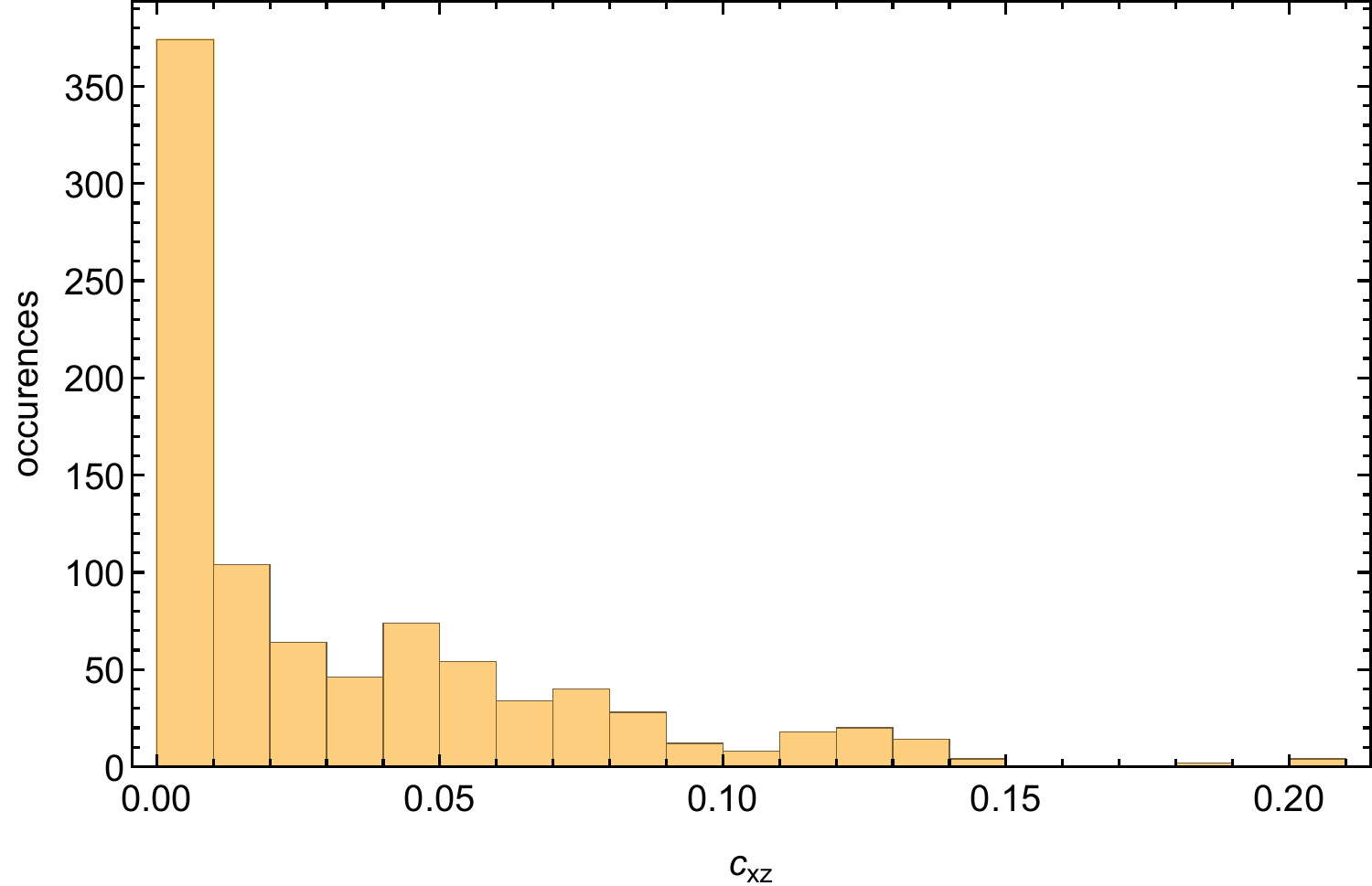}
	\caption{Histogram of the individual overlap elements $c_{xz}$ for $L = 30$ lattice sites at tunneling time $t = 0.5 L$.}
	\label{f:disting:overlap-histogram}
\end{figure}

Figure \ref{f:disting:overlap-max} shows the logarithm of the maximal value in this matrix as an indication of how close this pair of measurements comes to being mutually unbiased. 
For two lattice sites, given the correct tunneling time, a MUB is actually achievable, while for $L>2$ the two measurements are never maximally complementary. To implement two mutually unbiased measurements using  only position measurements after some tunneling evolution one would have to engineer tunnel couplings beyond nearest-neighbor tunneling, which is experimentally challenging.

Instead of looking only at the maximal overlap element, Fig.~\ref{f:disting:overlap-histogram} shows a histogram of all the elements of the overlap matrix for a special case. One finds that most elements are actually much smaller than the maximum, so using state-dependent bounds should give a significant benefit.

\subsubsection{Numerical Results}

Given that the state is close to a maximally entangled state, we expect correlations to be maximal when the two particles are measured in the same basis. Thus, in our entropic uncertainty relation we restrict ourselves to this case and evaluate $H(X_A|X_B)$ and $H(Z_A|Z_B)$.
Since the Schmidt basis of the ground state for attractive interactions is close to the site basis, the measured entropy $H(X_A|X_B)$ will turn out to be effectively zero in the following simulations. The measurable entanglement is thus determined by the relation between the correlations in our second measurement and by the complementarity factor $q$. 

\begin{figure}
	\centering
	\includegraphics[width=\linewidth]
	{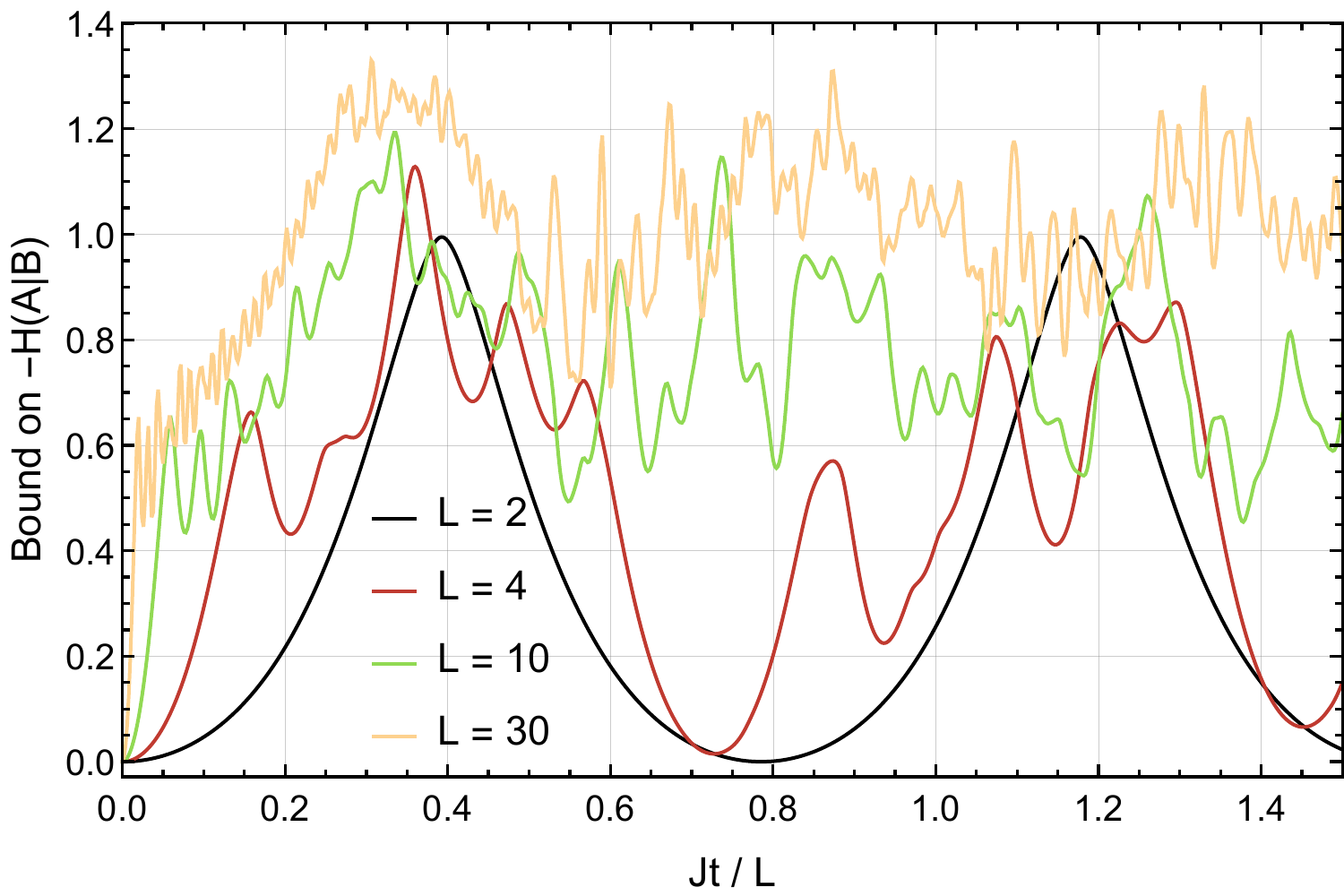}
	\caption{Detectable entanglement in the ground state for attractive interactions ($U/J = -100$) using the fully-state-dependent relation (\ref{r:eur_bipartite_fsdep}) and independent tunneling as the transformation to the second measurement basis. The abscissa shows tunneling time over the number of lattice sites, which parametrizes all such transformations. Even though the true entanglement entropy of the considered ground state grows as $-\qent(A|B) \approx \log(L)$, the detectable entanglement using this method is roughly constant.}
	\label{f:disting:tunneling-bounds}
\end{figure}
\begin{figure}
	\centering
	\includegraphics[width=\linewidth]{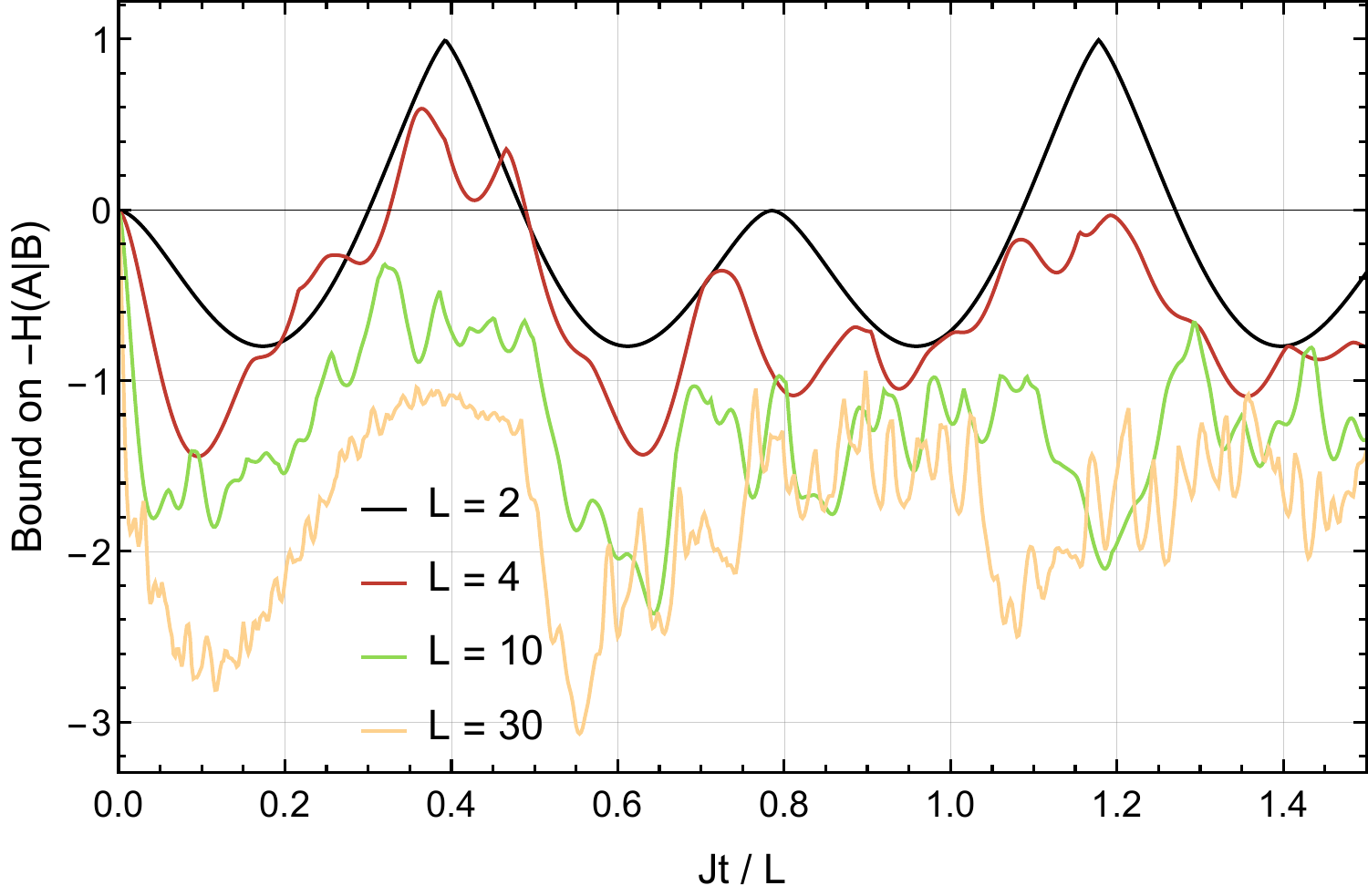}
	\caption{Similar to Fig.~\ref{f:disting:tunneling-bounds} but using the state-independent uncertainty relation (\ref{r:eur_bipartite_mu}). The detectable entanglement is significantly lower and for high number of sites no entanglement can be detected.}
	\label{f:disting:tunneling-bounds-stindep}
\end{figure}

The detectable entanglement using our fully-state-dependent relation (\ref{r:eur_bipartite_fsdep}) and the state-independent relation (\ref{r:eur_bipartite_mu}) is shown in Figs.~\ref{f:disting:tunneling-bounds} and \ref{f:disting:tunneling-bounds-stindep}, respectively. 
It is immediately apparent, that for everything which is not very close to a MUB measurement (this includes all measurements with high number of sites) the state-independent relation never detects any entanglement. Only for the special case of two sites, a tight quantification can be achieved using the state-independent bound. 
Our fully-state-dependent relation detects entanglement for all possible lattice configurations and tunneling times (even when the second measurement is very close to the first one). However, the detected entanglement seems to be limited to around 1.5 bits, whereas the true entanglement grows approximately as $\log(L)$. Thus, even though we can detect entanglement for all system sizes and chosen bases (tunneling times), using this method we cannot quantify entanglement accurately for many lattice sites in the sense that the obtained bounds are not tight.

Using the marginal-dependent relation (\ref{r:eur_coles_improved}) gives almost no benefit compared with the state-independent relation, as the row-maxima of the overlap matrices $\max_x c_{xz}$ are all very close to the global maximum (data not shown).  

\subsection{Spin-1 BEC}

As a second system, we consider a spin-1 Bose-Einstein condensate that is initially prepared in one spatial mode and subsequently split into two parts, which make up the two components of our bipartite system. 

Such systems have recently been used to demonstrate bipartite entanglement using a steering bound related to the Robertson uncertainty relation \cite{kunkel_spatially_2018, fadel_spatial_2018, lange_entanglement_2018}. While an application of entropic uncertainty relations using the same readout scheme fails, using different measurements our fully-state-dependent relation can be used to obtain a bound on an entanglement quantifier. The following system is modeled closely after the experimental procedure in Ref.~\cite{kunkel_spatially_2018}. 

\subsubsection{Model}
All particles of a $^{87}$Rb BEC are described as occupying the same spatial mode. The relevant internal states of the atoms, or spin states, are the three Zeeman sublevels of the $F=1$ hyperfine manifold of the $5s$ electronic ground state, labeled by $(1,0,-1)$. Spin mixing dynamics in this system are described within the Hilbert space spanned by the Fock states $\ket{N_1, N_0, N_{-1}}$, labeled by the number of particles in each spin component. This system can thus also be described as three harmonic oscillators, with corresponding ladder operators
\begin{equation} 
a_{-1}, a_0, a_1, \qquad a^\dagger_{-1}, a^\dagger_0, a^\dagger_1 \,.
\end{equation}
These three different modes correspond to the spin components in what we call the $z$-direction, i.e. 
\begin{equation} 
S_z = \hat{N}_1 - \hat{N}_{-1} = a^\dagger_1 a_1 - a^\dagger_{-1} a_{-1} \,.
\end{equation}

We call the basis constructed out of the occupation number states $\ket{N_1, N_0, N_{-1}}$ the bare mode basis. Experimentally, one can measure in this basis by applying a Stern-Gerlach pulse to separate the three spin components and then measure the number of particles in each mode by absorption imaging. The set of possible outcomes $x$ of a measurement will thus consist of all possible mode occupations $(N_1, N_0, N_{-1})$ consistent with $\sum_i N_i = N$, where $N$ is the total particle number. The probability distributions $P_X(x) = P(N_1, N_0, N_{-1})$, or, in the bipartite case discussed below, the joint distributions over the outcomes in the local subsystems, are basic quantities that need to be measured in order to obtain the conditional entropies $H(X_A|X_B)$ used in the entropic uncertainty relations.

The splitting into two subsystems is performed experimentally by letting the system expand in space and then measuring with spatial resolution, which allows us to split the measured absorption signal into two spatial parts. 
We model this splitting by moving to a bipartite Hilbert space $\HS = \HS_A \otimes \HS_B$ with three spin modes in each subsystem. 
This gives six total modes with corresponding operators
\begin{equation}
a_{A, -1},\, a_{A, 0},\, a_{A, 1}, \qquad a_{B, -1}\, , a_{B, 0}\, , a_{B, 1}
\end{equation}
and similar for $a^\dagger$. 
The transition from the single spatial mode to the two subsystems $A$ and $B$ can then be understood as replacing
\begin{equation}
a^\dagger_k \rightarrow \frac{1}{\sqrt{2}}\left( a^{\dagger}_{A, k} + a^{\dagger}_{B, k} \right) \,,
\label{bec:beamsplitting_definition}
\end{equation}
i.e., each particle has equal probability to end up in one of the two subsystems. This is equivalent to the application of a beam splitter to each mode.

\subsubsection{Measurement in Different Bases}

Measurements in bases other than the bare mode basis can be realized by time evolving with some Hamiltonian before measuring. Effectively, we apply the unitary rotation
\begin{equation}
R = \exp(-i t H) \,.
\end{equation}
Here we consider local spin rotations, i.e., Hamiltonians that consists only of pairs of creation and annihilation operators, so
\begin{equation} \label{spin1:su3_algebra}
R = \exp(i \sum_{j, k} C_{jk} a^\dagger_j a_k) \,.
\end{equation}
These form a representation of $\gU(3)$ and can be implemented experimentally by using external driving fields \cite{stamper-kurn_spinor_2013, kunkel_spatially_2018, kunkel_splitting_2019, Hamley2012}. 

Since changing a state $\ket{\psi}$ only by a global phase has no physical effect, for every $\gU(3)$ element there is a physically equivalent $\gSU(3)$ element. In the following we restrict ourselves to measurements which are related to the bare mode basis by such a $\gSU(3)$ transformation. This choice is a strong restriction in terms of possible choices of measurement bases but it appears reasonable from an experimental perspective, since interacting Hamiltonians are typically harder to engineer and control.
It will not include MUBs, but one can still ask which transformation $R$ on a Hilbert space with fixed particle number leads to a minimal $\max_{j,k} \mabs{R_{jk}}^2$.
It might be natural to start with $r \in \gSU(3)$ such that its fundamental representation on $\complex^3$ (which we write as $r$ again) has minimal $\max_{j,k} \mabs{r_{jk}}^2$.
The minimum of $\mabs{r_{jk}}^2 = \frac{1}{3}$ is achieved by the Fourier matrix
\begin{equation}
(F_3)_{jk} = \frac{i}{\sqrt{3}} \exp(\frac{2 \pi i j k}{3})\qquad j,k = 0, 1, 2\,,
\end{equation}
where we put a leading $i$ to make it an element of $\gSU(3)$.

We call the transformation associated with its representation
\begin{equation}
R_{FT} = \exp(\sum_{j, k} \log(F_3)_{jk} a^\dagger_j a_k) 
\end{equation}
the single-particle Fourier transformation, because it would be the Fourier transformation for only a single particle. 

In the regime where we can calculate representation matrices $R$ easy enough so that numerically optimizing over all of $\gSU(3)$ is possible, the single-particle Fourier transformation indeed appears to be optimal, i.e., there appears to be no other representation element $R'$ with smaller $\max_{j,k} \mabs{R'_{jk}}^2$. An analytical confirmation of this would be desirable, but is not straightforward to obtain.

\subsubsection{Configurational and Particle Number Entanglement}

Since any such $\gSU(3)$ operations preserve the particle number in each subsystem, Theorem \ref{theorem:conserved_quantities} applies. As mentioned in the previously shown example, for pure states with this setup we have
\begin{equation}
\begin{split}
- \qent{A|B} &= \qent{ρ_{B}} \\
&= \ent(\{p(n)\}) + \sum_n p(n) \qent{ρ_B^{(n)}} \\
&= \ent(\{p(n)\}) - \qent{A|B}_{\rb} \,,
\end{split}
\end{equation}
where $p(n)$ is the probability distribution of particle number $n$ in subsystem $A$ (or $B$). 
We call the first term $\ent(\{p(n)\})$ particle number entanglement (or particle number contribution to the entanglement entropy), and the second term $\sum_n p(n) \qent{ρ_B{(n)}}$ configurational entanglement (or configurational contribution to the entanglement entropy), in accordance with Ref.~\cite{Lukin2019}. Our uncertainty relations can only detect configurational entanglement, so we will always compare our bounds on $-\qent(A|B)$ with the configurational part of the true entanglement entropy. Note that the particle-number contribution does not actually depend on the details of the state but only on the total particle number. 

Ways to circumvent this undetectability of parts of the entanglement entropy likely go through circumventing the conservation law itself, for example, by interfering multiple copies of the state together \cite{islam_measuring_2015}.

\subsubsection{Overlaps on one Subsystem}
The complementarity factor $q$ used in entropic uncertainty relations quantifies the complementarity of the two measurements on the subsystem $A$. 
For this setup, the fact that the particle number in the subsystem is not fixed leads to some peculiarities when applying entropic uncertainty relations. 

Since the particle number in $A$ is not fixed, $\HS_A$ decomposes into subspaces of particle number $n$ in $A$:
\begin{equation}
\HS_A = \bigoplus_{n = 0}^N \HS^{(n)} \,.
\end{equation}
A representation $R$ of a $\gSU(3)$ element will then act on $\HS_A$ as
\begin{equation}
R_A = \bigoplus_{n = 0}^N R^{(n)} \,,
\end{equation}
where $R^{(n)}$ acts on $\HS^{(n)}$. 
To calculate the maximal matrix element we maximize over $n$ as well, so
\begin{equation}
\max_{j,k} \abs{(R_A)_{jk}} = \max_n \max_{j,k} \abs{\left(R^{(n)}\right)_{jk}} \,.
\end{equation}
However, since $R^{(0)} = \begin{pmatrix} 1 \end{pmatrix} \in \complex^{1 \cross 1}$ we have $\max_{j,k} \mabs{(R_A)_{jk}} = 1$ and thus $q_{MU} = - \log(1^2) = 0$. 

Thus, the state-independent relation (\ref{r:eur_bipartite_mu}) will never witness any entanglement for states on Hilbert spaces that allow for particle number fluctuation between subsystems. The key issue is that the beamsplitter allows for the possibility of all particles ending up in subsystem $B$. Since the overlap element of the uncertainty relation is state-independent and does not know anything about the beamsplitter, it has to acknowledge that there are states for which subsystem $A$ is empty. For these states, the measurement in $A$ is deterministic irrespective of any spin rotations applied, so there cannot be a nontrivial uncertainty relation. 

Thus, there has to be some state-dependence in the complementarity factor $q$. 
This can be achieved in various ways with increasing degree of state-dependence. The simplest modification just includes the overlaps of the state with the particle number sectors $\HS^{(n)}$ of our Hilbert space $\HS_A$ (this is effectively the probability distribution $\PP[N_A]{n}$ of finding $n$ particles in subsystem $A$),
\begin{equation}\label{d:spin1:q_PN}
q_{PN} \coloneqq -\sum_n \PP[N_A]{n} \log(\max_{j,k} \abs{\left(R^{(n)}\right)_{jk}}^2) \,.
\end{equation}
A relation with this $q$ is a simple corollary of (\ref{r:eur_coles_improved}). As an alternative, we may also use (\ref{r:eur_coles_improved}) directly, which yields
\begin{equation}\label{d:spin1:q_C}
q_{C} = -\sum_x \PP[X]{x} \log(\max_z c_{xz})
\end{equation}
with $\PP[X]{x}$ being the probability of measuring outcome $x$ (of measurement $X$) in $A$, and $\coverlap{x}{z} = \mabs{(R_A)_{xz}}^2$ in the previous notation. Here we take into account not only the probability distribution of finding a certain particle number in $A$, but the whole marginal probability distribution in $A$ (where the measurement results also imply a certain particle number). Finally, we may use the fully-state-dependent relation (\ref{r:eur_bipartite_fsdep}), which becomes (after setting $X = Y$) 
\begin{equation}\label{d:spin1:q_FSD}
q_{FSD} = -\sum_{x, y} \PP[X{}X]{x_A, x_B} \log(\sum_z c_{xz}\/ \PP[Z{}X]{z_A| x_B}) \,.
\end{equation}
This takes into account the full bipartite probability distribution $\PP[X{}X]{x_A, x_B}$ and also requires knowledge of the probability distribution $P_{Z{}X}$, where we do a local rotation in one subsystem only. Setting $X = Y$ in this relation gives an entropic uncertainty relation which has $H(X_A|X_B)$ on its left-hand side. For this specific systems conditioning on the same measurement on the other subsystem always appears to be optimal. 
For $q_C$ and $q_{FSD}$ we can get a potentially different $q$ by swapping $X$ and $Z$. In practice we take the maximum of the two. 

For the purpose of entanglement quantification in experiment, where we need to use measured entropies $H(X_A|X'_B) + H(Z_A|Z'_B)$ on the left-hand side of our uncertainty relations, we seek to maximize the complementarity factor. Given that $q_{PN}\leq q_{C}\leq q_{FSD}$, the fully-state-dependent relation implies the other two relations.
We will see below that the fully-state-dependent relation is the only one which can realistically be used to certify entanglement in this setup. 

\subsubsection{Two-Mode-Squeezed State}
An interesting (and experimentally accessible) set of entangled states is formed by the so called two-mode-squeezed states. These are created from a prepared state $\ket{0, N, 0}$ by time evolution under a Hamiltonian that allows for spin-changing collisions \cite{Hamley2012, kunkel_spatially_2018, lange_entanglement_2018} 
\begin{multline}\label{spin1:hamiltonian_scc_full}
H = g\Bigg( a^\dagger_1 a^\dagger_{-1} a_0 a_0 +  a^\dagger_0a^\dagger_0 a_1 a_{-1}{}+{}  \\  \left(\hat{N}_0 - \frac{1}{2}\right)(\hat{N}_1 +\hat{N}_{-1}) \Bigg)  + q(\hat{N}_1 + \hat{N}_{-1}) \,.
\end{multline}
The ratio between the parameters $g$ and $q$ can be tuned in experiment. For the following sections on squeezed states we set $q = - g\,(N - \frac{1}{2})$ with $N$ the total particle number. This ensures that for high occupation of the zero mode $N_0 \approx N$ the $q$-dependent term cancels the second $g$-dependent term, and we are left with
\begin{equation}
H \approx g\left( a^\dagger_1 a^\dagger_{-1} a_0 a_0 +  a^\dagger_0a^\dagger_0 a_1 a_{-1} \right) \,.
\end{equation}

For short evolution times we can assume the zero-mode population to remain constant (undepleted pump approximation) and thus replace the corresponding operators by constant numbers, so 
$\hat{N}_0 \sim N$ and $a_0 \sim a^\dagger_0 \sim \sqrt{N}$. Then the Hamiltonian can be approximated as
\begin{equation}
H = g N(a^\dagger_1 a^\dagger_{-1} + a_1 a_{-1} ) \,,
\end{equation}
which is solvable analytically. 
Introducing the squeezing parameter $r = N{}g t$ one gets \cite{Braunstein2005, kunkel_spatially_2018}
\begin{equation}\label{spin1:udp_state}
\ket{ψ(t)} = \frac{1}{\cosh(r)} \sum_n (- i \tanh(r))^n \ket{n, N - 2n, n} \,.
\end{equation}
This holds only for small $r$, but can give good insight into the approximate structure of the state. For the following numerical computations we still diagonalize the full Hamiltonian (\ref{spin1:hamiltonian_scc_full}) and calculate the time evolution explicitly.  All subsequent results which depend on squeezing will be shown as a function of $r = N{}gt$. 

The reason this is called a squeezed state is that there exists a pair of observables with almost canonical commutation relation, so that for increased squeezing one becomes increasingly localized while the other one becomes increasingly delocalized (similar to squeezing of position and momentum).
With \begin{equation}\label{spin1:squeezed_spin_observable}
S(φ) = \frac{i}{\sqrt{2}}(e^{-i φ} a^\dagger_0(a_1 - a_{-1}) + h.c.)
\end{equation}
we define the spin the squeezed direction as $\spif$ and the spin in the antisqueezed direction as $\stpif$. 
Then, 
\begin{align}\label{spin1:sq_antisq_comm}
\left[\spif, \stpif\right] &=  2i  \hat{N}_{0} - i (a^\dagger_1 + a^\dagger_{-1})(a_1 + a_{-1}) \notag \\ &\sim 2 i N_0 \approx const.
\end{align}
for low squeezing \cite{kunkel_spatially_2018}. 
Alternatively one can also characterize these operators by the rotations which relate them to $S_z$.
These read
\begin{subequations}
\begin{align}
\label{spin1:squeezed_rotation}
R_{sq} &= e^{- i \frac{π}{4} \hat{N}_0}\, e^{-i \frac{π}{2} \hat{S}_y} \\
\label{spin1:antisqueezed_rotation}
R_{antisq} &= e^{- i \frac{3π}{4} \hat{N}_0}\, e^{-i \frac{π}{2} \hat{S}_y} \,.
\end{align}
\end{subequations}

\subsubsection{Numerical Results: Time-evolved State}

All numerical results for the spin-1 BEC consider the pure state prepared by unitary time evolution under Hamiltonian \eqref{spin1:hamiltonian_scc_full} with a single fixed total particle number. In practice this usually requires postselection of measurements. 

Previous demonstrations of entanglement in these systems \cite{kunkel_spatially_2018} used the squeezed and antisqueezed direction as measurement pairs, as the Robertson relation is tight for these measurements. Entropic uncertainty relations are 
not tight for these measurements however, and even our fully-state-dependent relation fails to witness any entanglement in this case. 

For entropic uncertainty relations we find different pairs of measurements by numerical optimization. In general we would like to optimize the fully-state-dependent relation over both measurement settings, i.e., over $\gSU(3) \cross \gSU(3)$. However, doing this for a particle number that is not too small to show similar behavior as the high $N$ limit is practically difficult. If we fix the single-particle Fourier transformation as the transformation between the two measurements we can minimize $H(X_A|X_B) + H(Z_A|Z_B)$ over the $\gSU(3)$ set of all measurements $X$. The optimal measurement then depends on the state. For squeezed states we find two different optima at different values of $r$. 

\begin{figure}
	\includegraphics[width=\linewidth]{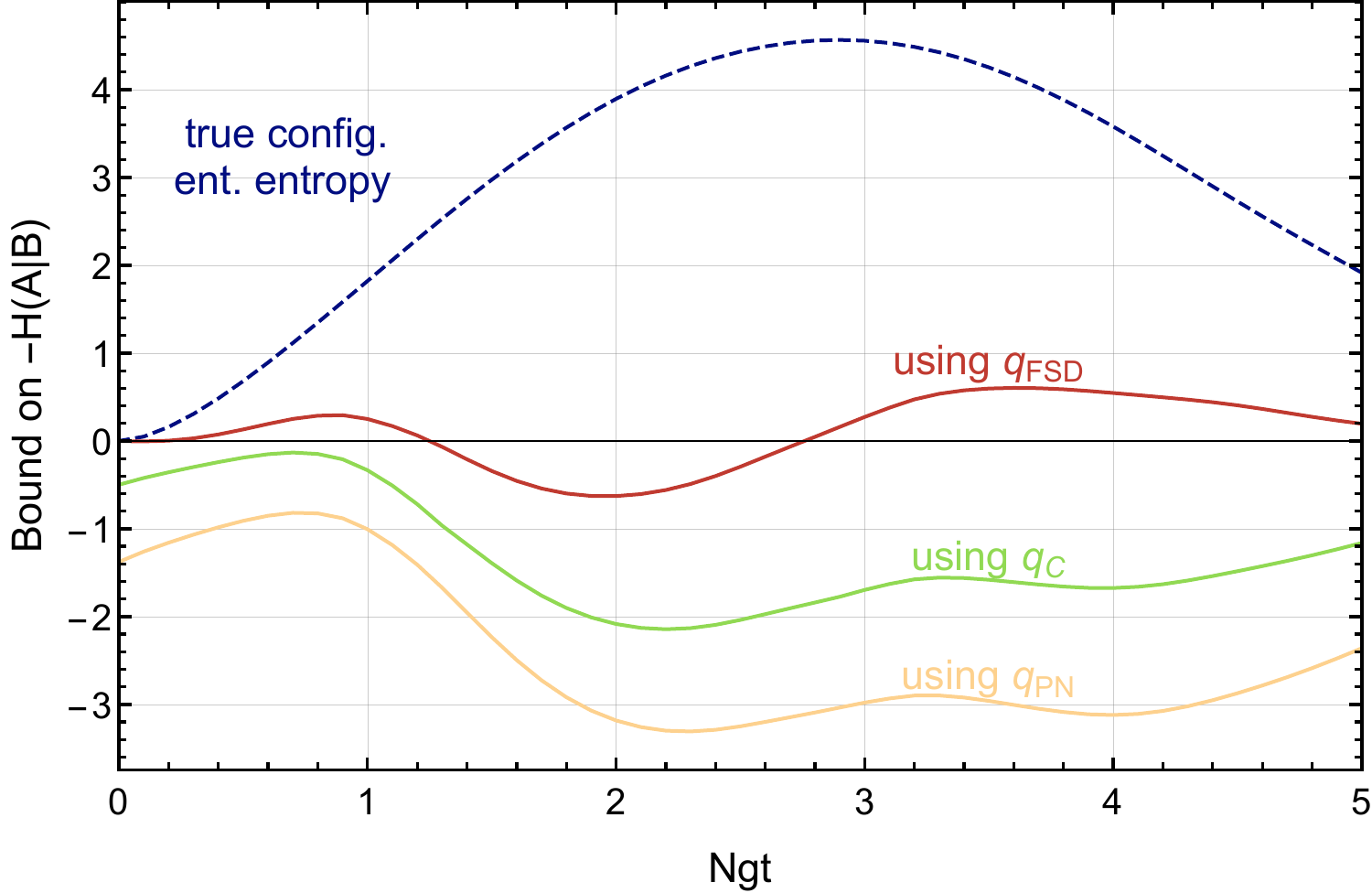}
	\caption{Detectable entanglement for the time-evolved state of a spin-1 BEC. Bounds on the entanglement entropy using a numerically optimized first measurement and its single-particle Fourier transform. Numerical optimization has been applied to maximize detection at $r = 0.5$ for $N = 15$. The optimal basis choice was used to produce the data for $N = 50$ shown in the figure.}
	\label{f:spin1:goptbound}
\end{figure}

In the short-time regime (at $r = 0.5$ for 15 particles) we find a measurement for which a small violation [i.e.,  $q > H(X_A|X_B) + H(Z_A|Z_B)$] exists, as shown in Fig.~\ref{f:spin1:goptbound}. 
In terms of parametrizations of $\gSU(3)$, the numerical minimum is not unique. However, all minimizers seem to be only rotating individual (spin-)modes by a complex phase and do not perform any mode mixing. 
Since this initial rotation actually does not influence the first measurement results in of itself, it should be understood as preparation for making the application of the Fourier transformation more effective, and thus could also be included into the transformation between the two measurement directions, which would then be a slightly modified Fourier transformation. 

The minimum is attained for the three complex phases
\begin{equation}
(\phi_1, \phi_0, \phi_{-1}) = (0.095 \pi, -0.495 \pi, 0.400 \pi)
\end{equation}
which add up to zero since we are implementing a $\gSU(3)$ rotation. 

If we restrict ourselves to performing only such a $\gU(1)\,\cross\, \gU(1)$ rotation for the first measurement direction, we can numerically minimize not only $H(X_A|X_B) + H(Z_A|Z_B)$ but the bound on $-H(A|B)$ through the fully-state-dependent entropic uncertainty relation. It converges to the same minimum as achieved by just optimizing $ H(X_A|X_B) + H(Z_A|Z_B)$.

\begin{figure}
	\includegraphics[width=\linewidth]{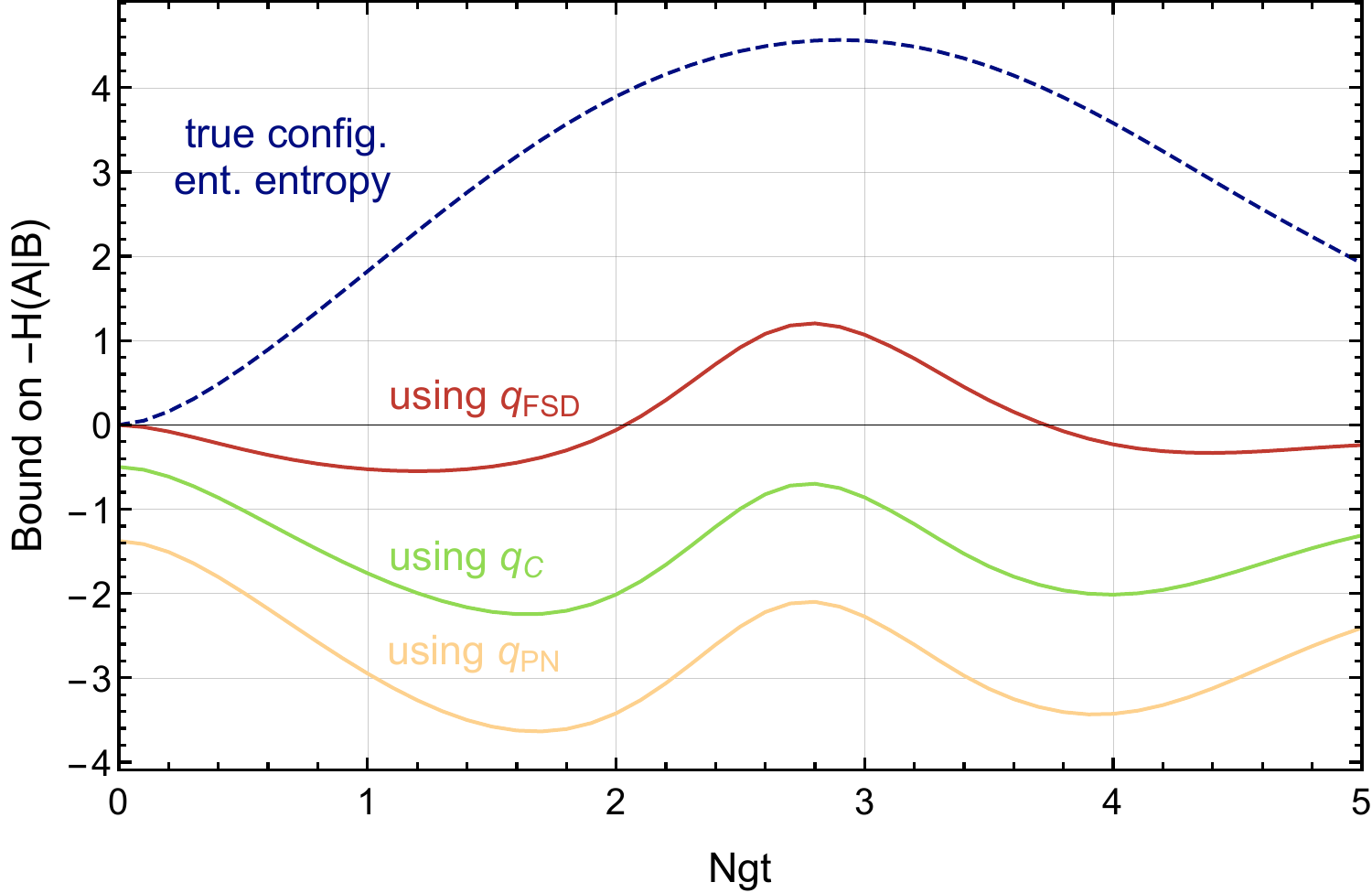}
	\caption{Bounds on the entanglement entropy using a measurement in the bare mode basis and its single-particle Fourier transformation. The prepared state is the same as in Fig.~\ref{f:spin1:goptbound}.}
	\label{f:spin1:bmftbound}
\end{figure}

Doing the numerical optimization in the over-squeezed (long-time) regime (at $r = 2.5$ for 15 particles), it appears that using the bare mode basis as the first measurement direction is almost optimal. Slight improvements exists, but they do seem to depend on the particle number and do not show any overall different behavior. The bare mode basis and its single-particle Fourier transform are shown in Fig.~\ref{f:spin1:bmftbound}. At the maximum point in the over-squeezed regime, we can witness a significant amount of entanglement. 
 
Numerical optimizations at squeezing parameters $r$ in between the above two values always seem to converge to either one of the two results shown above, or something which never witnesses any entanglement. 
In any case, only the fully-state-dependent relation is tight for the initial state (which contains no configurational entanglement) and manages to certify some entanglement for $r>0$.

\subsubsection{Numerical Results: Ground States}

Besides time-evolved states, we can also look at ground states of the Hamiltonian (\ref{spin1:hamiltonian_scc_full}). 
We will only consider ferromagnetic condensates, for which $g < 0$, as realized for the $F = 1$ hyperfine manifold of $\,\prescript{87}{}{\mathrm{Rb}}$ \cite{stamper-kurn_spinor_2013}. In this case, in the limit $N \to \infty$, the system shows two second-order phase transitions at 
\begin{equation}\label{d:spin1:gs:q_critical}
q = \pm q_c \qquad q_c = 2 N \abs{g}
\end{equation} \cite{stamper-kurn_spinor_2013, zhang_generation_2013, feldmann_interferometric_2018}. For $q > q_c$ the system is in the polar phase with the polar ground-state $\ket{\psi_p} = \ket{0, N, 0}$. This state is separable between the individual spins and will possess only particle-number entanglement when split into two spaical parts.  For $q < -q_c$ the system is in the TF phase with ground state $\ket{\psi_{TF}} = \ket{\frac{N}{2}, 0, \frac{N}{2}}$, or $\ket{\psi_{TF}} = \ket{\lfloor\frac{N}{2}\rfloor, 1, \lfloor\frac{N}{2}\rfloor}$ if $N$ odd. For small $N$, this difference between $N\/$ odd and even has a significant impact on the entanglement entropy.
For $\abs{q} \leq q_c$ the ground state occupies modes $\ket{k, N-2k, k}$, with $k$ transitioning from $0$ to $N{}/2$ as $q$ decreases. 

\begin{figure}
	\includegraphics[width=\linewidth]{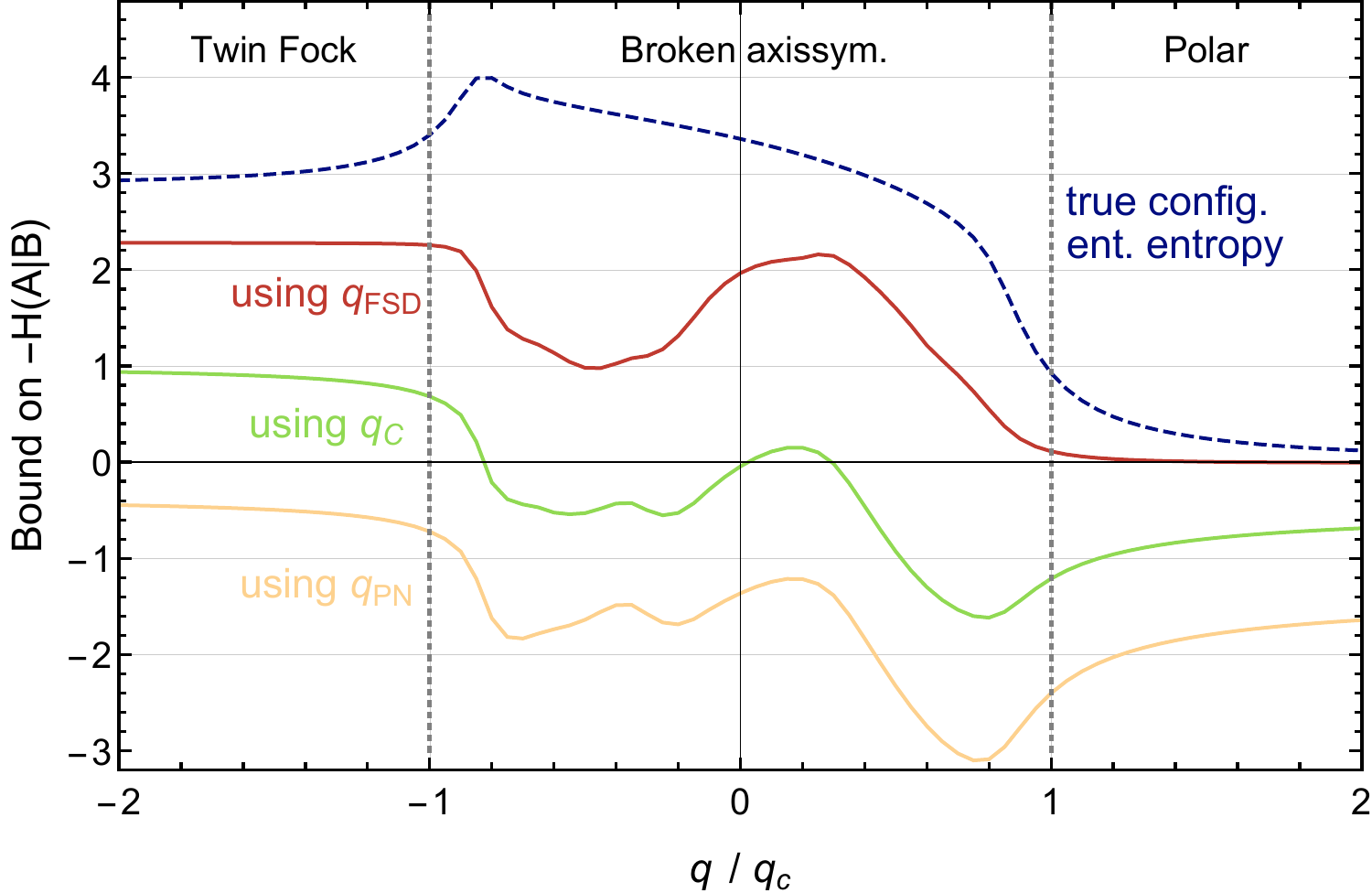}
	\caption{Bounds on $-\qent(A|B)$ in the ground state of the spin-1 BEC using multiple complementarity factors as a function of the Hamiltonian parameter $q$ for $N = 50$ particles. Measurements are performed in the bare mode basis and its single-particle Fourier transform.}
	\label{f:spin1:gs_bounds}
\end{figure}

As in the case of the time-evolved states, we use measurements in the bare mode basis and the single-particle Fourier-transformed basis.
For these two measurement directions, Fig.~\ref{f:spin1:gs_bounds} shows $-H(A|B)$ and bounds on it through entropic uncertainty relations as a function of the Hamiltonian parameter $q$.
The entanglement entropy shows a significant increase at the $q = q_c$ phase-transition point, and also the TF-ground state shows significant configurational entanglement. 

Similar to what we saw for squeezed states, only the fully-state-dependent relation is tight for the (configurationally separable) polar state and can properly detect the increase in entanglement at the phase transition point. Also in the TF ground state the fully-state-dependent relation yields a significantly larger amount of certified entanglement.

\section{Conclusions and Outlook}\label{sec:conclusion}

We derived an improved entropic uncertainty relation that allows the quantification of entanglement based on measurements in only two different local measurement settings, which need not be mutually unbiased. As proof of principle we demonstrated its use through two numerical studies of model systems inspired by previously performed experiments. Our method can potentially be implemented on any experimental platform, provided the possibility of single-particle resolved detection and measurements in two different suitable local bases. These requirements are met by most of the currently available quantum simulation platforms ranging from cold atoms in optical lattices \cite{Kuhr2016} or tweezer arrays \cite{Browaeys2020} to trapped ions \cite{Blatt2012} and superconducting qubits \cite{Kjaergaard2019}. 
Entropic entanglement bounds may thus find applications ranging from the quantum simulation of quantum statistical mechanics problems \cite{Schreiber2015, Smith2016} to high-energy physics \cite{Banuls2020}, where quantifying entanglement is crucial for addressing fundamental questions.

A topic that will require further investigation when applying our method in experiments is the role of noise and finite measurement statistics.
In our example applications we considered pure states that can be measured with arbitrary precision, so that the full probability distribution over the possible measurement outcomes is available. Realistic experimental realizations have to deal with preparation and readout noise and finite measurement statistics and can thus estimate the conditional entropies only up to some experimental error. The accuracy of entropy estimation in the presence of such errors will in general depend on the state in question, and it is unclear a priori how for some given situation a worst-case estimate would look like. Also, the tightness of our entanglement bounds may depend on the purity of the prepared states.

This leads to the question of scalability of our approach to entanglement quantification. In the worst case, an accurate estimate of the entropy requires $O(\dim \HS)$ measurement samples, but this can be significantly improved if the probability distribution is highly localized \cite{schneeloch_quantifying_2019}. Finding out to which degree such localized distributions can always be found and used for such a procedure is another question to be addressed. Also, in general, the complementarity factors need to be calculated numerically which scales at least as $O(\dim \HS)$. Analytical expressions are only available for specific basis choices such as MUBs. It will thus be important to obtain analytical results for larger classes of bases pairs allowing for an efficient evaluation of complementarity factors or at least their asymptotic behavior at large systems sizes.

\acknowledgements
We thank Stefan Flörchinger, Tobias Haas, Philipp Kunkel, and Markus Oberthaler for helpful discussions. This work is supported by the Deutsche Forschungsgemeinschaft (DFG, German Research Foundation) under Germany’s Excellence Strategy EXC2181/1-390900948 (the Heidelberg STRUCTURES Excellence Cluster) and within the Collaborative Research Center SFB1225 (ISOQUANT).

\appendix*
\section{Zero Quantum Discord States}
    Let $X$ be an orthonormal basis of $\HS_B$ consisting of states $\ket{x}$, and $ρ_{AB}$ be a density matrix on $\HS_A \otimes \HS_B$. We write $ρ_{AX}$ for the state obtained after measuring subsystem $B$ in the $X$ basis, i.e.,
    \begin{equation}
        ρ_{AX} = \sum_x (\IdentityMatrix_A \otimes \ketbra{x}{x}) ρ_{AB}  (\IdentityMatrix_A \otimes \ketbra{x}{x})\, ,
    \end{equation}
    and $H(A|X)$ for the conditional entropy of this state. We say the state $ρ_{AB}$ has zero quantum discord, if there exists an orthonormal basis $X$, such that
    \begin{equation}
        H(A|X) = H(A|B)\,.
    \end{equation}
\begin{lemma}[Zero Discord States]
\label{appendix:lemma_zero_discord}
A state $ρ_{AB}$ has zero quantum discord if and only if it is quantum-classical, i.e., taking $Z$ as the orthonormal basis in which $ρ_{B}$ is diagonal, it holds that
\begin{equation}
    ρ_{AB} = ρ_{AZ}\,.
\end{equation}
\begin{proof}
It is obvious, that $ρ_{AB}$ being quantum-classical implies zero quantum discord. For the other direction we follow the proof in Ref.~\cite{datta_condition_2011}. 

Let $X$ be an ONB such that $H(A|X) = H(A|B)$. We introduce a second copy of the Hilbert space $\HS_B$, which we call $\HS_C$, and denote by $V$ the isometry that acts as
\begin{subequations}
\begin{align}
    V&\colon \HS_B \to \HS_B \otimes \HS_C \\
    V&\ket{\psi} = \sum_x \ketbra{x}\ket{\psi} \otimes \ket{x}\,.
\end{align}
\end{subequations}
This corresponds to the isometry that implements the POVM of measuring in $X$. Similarly to what was done in Sec.~\ref{sec:povms}, we write $\tilde{ρ}_{AXX'} = Vρ_{AB}V^\dagger$, and get $\tilde{ρ}_{AX} = \tilde{ρ}_{AX'} = ρ_{AX}$, $\tilde{ρ}_X = \tilde{ρ}_{X'} = ρ_{X}$, as well as $H(\tilde{ρ}_{AXX'}) = H(ρ_{AB})$ and $H(\tilde{ρ}_{XX'}) = H(ρ_B)$ due to invariance under isometries. The equality $H(A|X) = H(A|B)$ then corresponds to equality in the strong-subadditivity relation:
\begin{equation}
    H(\tilde{ρ}_{AXX'}) + H(\tilde{ρ}_{X}) \leq H(\tilde{ρ}_{AX}) + H(\tilde{ρ}_{XX'}) \,.
\end{equation}

In Ref.~\cite{hayden_structure_2004} it was shown that a state $ρ_{ABC}$ satisfies equality in strong subadditivity if and only if there exists a decomposition of the Hilbert space $\HS_B$
\begin{equation} \label{appendix:orthogonal_support}
    \HS_B = \bigoplus_j \HS_{B_L^j} \otimes \HS_{B_R^j}
\end{equation}
such that
\begin{equation}
    ρ_{ABC} = \bigoplus_j p_j ρ_{A{B_L^j}} \otimes ρ_{{B_R^j}C} \, .
\end{equation}
Note that we can embed $\HS_B$ into 
\begin{equation}
\HS_{B_L} \otimes \HS_{B_R} \coloneqq \left(\bigoplus_j \HS_{B_L^j}\right) \otimes \left(\bigoplus_j \HS_{B_R^j}\right)
\end{equation}
and thus also write

\begin{equation}
    ρ_{ABC} = \bigoplus_j p_j ρ_{A{B_L}}^j \otimes ρ_{{B_R}C}^j \,,
\end{equation}
where the $ρ_{A{B_L}}^j$ have support only in $\HS_A \otimes \HS_{B_L^j}$ and similarly for $ρ_{B_R C}^j$.

We apply this result to the tripartite state $\tilde{ρ}_{AXX'}$ which achieves equality in strong subadditivity. This implies that $\HS_B$ has the given decomposition, and since $\HS_B = \HS_C$ the same holds for $\HS_C$. Note that the state $\rt_{AXX'}$ is fully symmetric under exchanging $X \leftrightarrow X'$. Let us consider the objects
\begin{equation}
\label{appendix:j-decomp}
    \rt_{AX_L X_R X'_L X'_R}^j = \rt_{AX_L}^j \otimes \rt_{X_R X'_L X'_R}^j
\end{equation}
which, according to the theorem just stated, sum to
\begin{equation}
    \rt_{AXX'} = \bigoplus_j p_j \rt_{AX_L X_R X'_L X'_R}^j \, .
\end{equation}
The exchange symmetry implies
\begin{equation}
    \rt_{AX_L}^j = \rt_{AX'_L}^j \quad \Rightarrow \quad \tilde{ρ}_{AX_L}^j = \rt_A^j \otimes \rt_{X'_L}^j \,,
\end{equation}
where we took partial traces of (\ref{appendix:j-decomp}), and thus we get
\begin{equation}
    \rt_{AXX'} = \bigoplus_j p_j \rt_A^j \otimes \rt_{XX'}^j \,,
\end{equation}
with $\rt_{XX'}^j = \rt_{X_L}^j\otimes \rt_{X_R X'}^j$. Now, (\ref{appendix:orthogonal_support}) states that all the $\rt_{XX'}^j$ have orthogonal support. 
We can invert the isometry $V$ to get
\begin{equation}
    ρ_{AB} = \sum_j p_j ρ_{A}^j \otimes ρ_{B}^j \,,
\end{equation}
where all the $ρ_{B}^j$ then have orthogonal support, and so the eigenbasis $Z$ of $ρ_{B}$ can be constructed out of eigenvectors of the $ρ_{B}^j$. Thus all the $ρ_{B}^j$ are diagonal in $Z$, which implies $ρ_{AB} = ρ_{AZ}$.

\end{proof}
\end{lemma}
\bibliography{refs}

\end{document}